%% file: SMP-21-001_temp.tex
\pdfoutput=1
\documentclass[11pt,twoside,a4paper,cmspaper,final,collab]{cms-tdr}
\def\svnVersion{f707143}\def\svnDate{2023/05/11}\def\cmsCernNoTag{CERN-EP-2022-038}\def\cmsCernDate{\today}\def\cmsMessage{Published in Physics Letters B as \href{http://dx.doi.org/10.1016/j.physletb.2022.137495}{\doi{10.1016/j.physletb.2022.137495}.}}
\begin{document}\cmsNoteHeader{SMP-21-001}

\newcommand{\ww}{\ensuremath{\PWp\PWm}\xspace}
\newcommand{\tautau}{\ensuremath{\PGtp{}\PGtm}\xspace}
\newcommand{\wj}{\ensuremath{\PW+\mathrm{jets}}\xspace}
\newcommand{\sqrts}{\ensuremath{\sqrt{s}}\xspace}
\newcommand{\sobs}{\ensuremath{\sigma^{\text{obs}}}\xspace}
\newcommand{\ssm}{\ensuremath{\sigma^{\text{SM}}}\xspace}
\newcommand{\detajj}{\ensuremath{\abs{\Delta\eta_{\mathrm{jj}}}}\xspace}
\newcommand{\mjj}{\ensuremath{m_{\mathrm{jj}}}\xspace}
\newcommand{\mt}{\ensuremath{m_{\mathrm{T}}}\xspace}
\newcommand{\mll}{\ensuremath{m_{\ell\ell}}\xspace}
\newcommand{\ptll}{\ensuremath{p_{\text{T}}^{\ell\ell}}\xspace}
\newcommand{\ptlone}{\ensuremath{\pt^{\ell_{1}}}\xspace}
\newcommand{\ptltwo}{\ensuremath{\pt^{\ell_{2}}}\xspace}
\newcommand{\ptvecll}{\ensuremath{\ptvec^{\ell\ell}}\xspace}
\newcommand{\Zll}{\ensuremath{Z_{\ell\ell}}\xspace}
\newcommand{\Zl}{\ensuremath{Z_{\ell}}\xspace}
\newcommand{\Zlone}{\ensuremath{Z_{\ell_{1}}}\xspace}
\newcommand{\Zltwo}{\ensuremath{Z_{\ell_{2}}}\xspace}
\newcommand{\etalep}{\ensuremath{\eta_{\ell}}\xspace}
\newcommand{\etajone}{\ensuremath{\eta_{\mathrm{j}_{1}}}\xspace}
\newcommand{\etajtwo}{\ensuremath{\eta_{\mathrm{j}_{2}}}\xspace}
\providecommand{\PVpr}{\HepParticle{V}{}{\prime}\xspace}
\providecommand{\PQqpr}{\HepParticle{q}{}{\prime}\xspace}
\ifthenelse{\boolean{cms@external}}{\newcommand{\cmsTable}[1]{\resizebox{\columnwidth}{!}{#1}}}{\newcommand{\cmsTable}[1]{#1}}
\ifthenelse{\boolean{cms@external}}{\providecommand{\suppMaterial}{the supplemental material [URL will be inserted by publisher]}}{\providecommand{\suppMaterial}{App.~\ref{app:suppMat}}}

\cmsNoteHeader{SMP-21-001}
\title{
Observation of electroweak \texorpdfstring{$\PW^{+}\PW^{-}$}{W+W-} pair production in association with two jets in proton-proton collisions at \texorpdfstring{$\sqrt{s} = 13\TeV$}{sqrt(s) = 13 TeV} 
}

\date{\today}

\abstract{An observation is reported of the electroweak production of a $\PWp\PWm$ pair in association with two jets, with both \PW bosons 
decaying leptonically. The data sample corresponds to an integrated luminosity of 138\fbinv of proton-proton collisions at $\sqrt{s}=13\TeV$, 
collected by the CMS detector at the CERN LHC. Events are selected by requiring exactly two opposite-sign leptons (electrons or muons) and two 
jets with large pseudorapidity separation and high dijet invariant mass. Events are categorized based on the flavor of the final-state leptons. 
A signal is observed with a significance of 5.6 standard deviations (5.2 expected) with respect to the background-only hypothesis. The measured 
fiducial cross section is $10.2\pm2.0\unit{fb}$ and this value is consistent with the standard model prediction of $9.1\pm0.6\unit{fb}$.}

\hypersetup{
pdfauthor={CMS Collaboration},
pdftitle={Observation of electroweak W+W- pair production in association with two jets in proton-proton collisions at sqrt{s} = 13 TeV},
pdfsubject={CMS},
pdfkeywords={CMS, W pairs, VBS}}

\maketitle

\section{Introduction}
\label{sec:introduction}
Electroweak (EW) scattering processes of the form $\PV\PVpr\to\PV\PVpr$, 
with \PV and \PVpr being \PW, \PZ, or \PGg vector bosons, are crucial tools to investigate the mechanism of EW symmetry breaking~\cite{vbstheory1,vbstheory2}. 
The existence of a Higgs boson with a mass of 125\GeV~\cite{atlashiggs2012,cmshiggs2012,cmshiggs2013} prevents the violation of unitarity in such vector boson 
scattering (VBS) processes by adding new exchange diagrams that cancel divergences in the theoretical calculations involving massive gauge bosons~\cite{PhysRevD.16.1519}. 
Therefore, the precise measurement of VBS cross sections in proton-proton ($\Pp\Pp$) collisions at the CERN LHC 
can probe the nature of the Higgs sector and search for effects beyond the standard model.

The cross sections of the various VBS processes and their associated backgrounds differ considerably 
depending on the choice of the pair of bosons and their final state.
The EW production of two W bosons with the same electric charge ($\PW^{\pm}\PW^{\pm}$) in the fully leptonic final state 
has been extensively studied by the ATLAS and CMS Collaborations~\cite{ATLAS8TeV, SMP-13-015, ATLAS2016, SMP-17-004, SMP-19-012, SMP-20-006}.
In this paper, the full 2016--2018 data set recorded by the CMS experiment is exploited to search for the 
purely EW production of a pair of opposite-sign (OS) \PW bosons in association with two jets, a process that 
has not been observed yet. 

This analysis, however, is more challenging than the $\PW^{\pm}\PW^{\pm}$ channel, mainly due to the large OS background 
from top pair (\ttbar) production, which results in a lower experimental sensitivity. From a theoretical perspective, 
they are both key ingredients to provide a full picture of the EW symmetry breaking mechanism in the SM, but the EW 
production of \ww bosons is more sensitive to new physics phenomena that might affect the couplings of \PW bosons to 
the Higgs boson. In view of a future combination of VBS channels to set constraints on effective-field theory parameters, 
this process is particularly important for its capability of precisely determining several dimension-6 operators~\cite{Bellan:2021dcy}. 

\begin{figure*}[!hbt]
  \centering
  \includegraphics[width=0.3\textwidth]{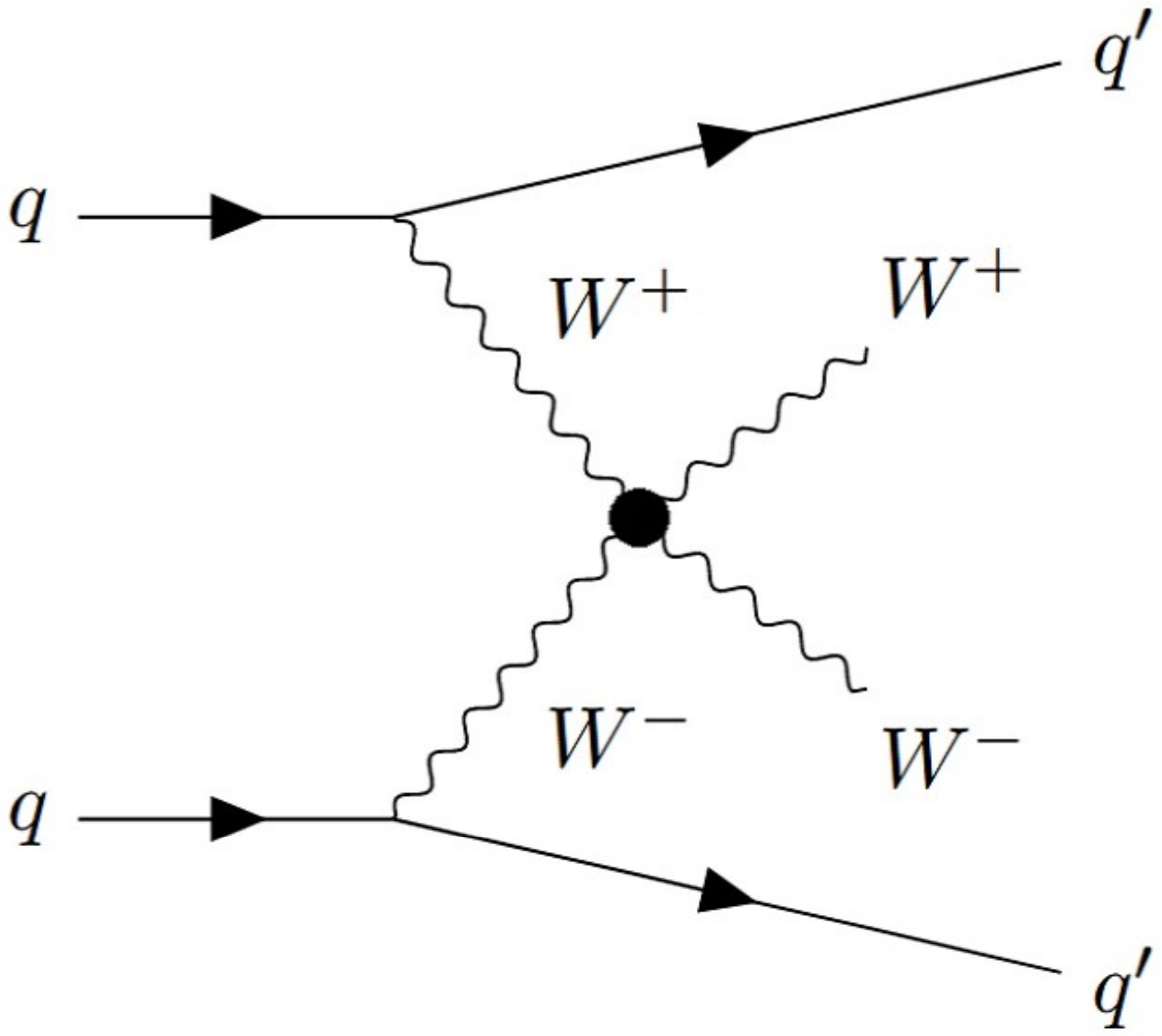}
  \includegraphics[width=0.31\textwidth]{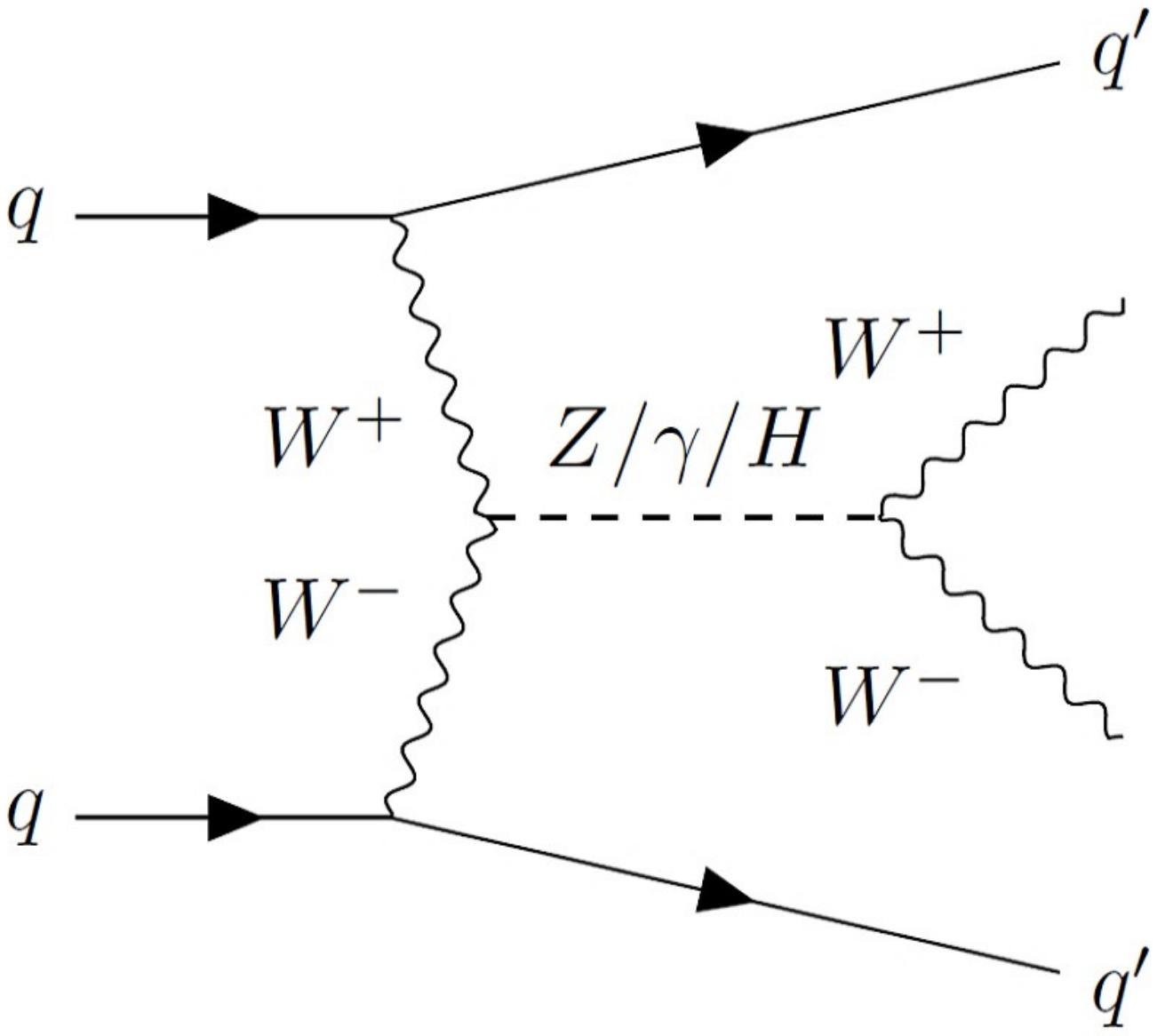}
  \includegraphics[width=0.3\textwidth]{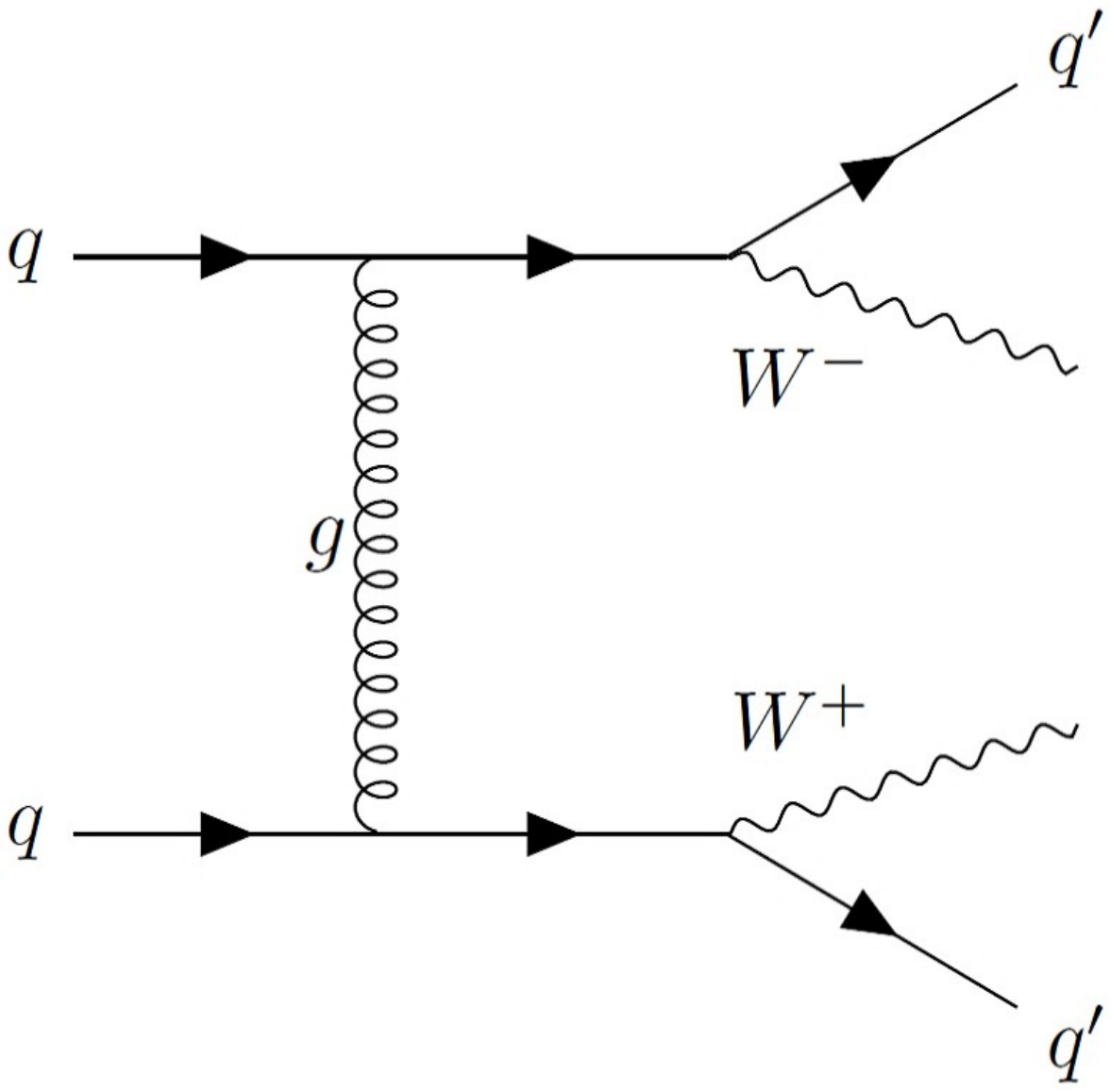}
    \caption{Examples of Feynman diagrams for the EW (left, center) and QCD-induced (right) production of \ww bosons in association with two quarks.}
  \label{fig:vbsos}
\end{figure*}

Fig.~\ref{fig:vbsos} (left and middle) shows some typical diagrams at the lowest order in the EW coupling (\ie, 
$\alpha^{6}$ including subsequent \PW boson decays) that contribute to the signal. In Fig.~\ref{fig:vbsos} (right), 
the production of the \ww final state is instead mediated via gluon exchange, referred to as quantum chromodynamics 
(QCD)-induced \ww production. These QCD diagrams constitute an irreducible background for the analysis, although they 
have different kinematic properties that can be used to reduce their contamination in the signal region (SR).

The characteristic signature of VBS events includes two vector bosons and two jets emitted at forward and backward rapidities.
The jets in the VBS topology (tagging jets) have large pseudorapidity separation (\detajj), a high dijet invariant mass (\mjj), and suppressed hadronic activity between them.
The analysis selects final states with two OS leptons ($\Pep\Pem$, $\PGmp\PGmm$, $\Pe^{\pm}\PGm^{\mp}$).
The $\Pe\Pgm$ channel is less populated by Drell--Yan (DY) events, thus resulting in a lower background contamination 
and higher sensitivity. The neutrinos in the final state result in large missing transverse momentum (\ptmiss).

The main source of reducible background arises from \ttbar production.  
This background is suppressed by vetoing the presence of jets originating from the fragmentation of bottom (\PQb) quarks, 
and its contamination in the SR is measured from CMS data through dedicated control regions (CRs) enriched in \ttbar events.
Together with the QCD-induced \ww production, which is reduced by requirements on \mjj and \detajj, 
other relevant sources of background events are DY plus jets and \PW boson plus jets production,
and their yields are all estimated from a CMS data set via dedicated CRs.

\section{The CMS detector}
\label{sec:CMS}

The central feature of the CMS apparatus is a superconducting solenoid of 6\unit{m} internal diameter, providing a magnetic field of 3.8\unit{T}. 
A silicon pixel and strip tracker, a lead tungstate crystal electromagnetic calorimeter (ECAL), and a brass and scintillator hadron calorimeter (HCAL), 
each composed of a barrel and two endcap sections, are installed within the solenoid. Forward calorimeters extend the $\eta$ coverage 
provided by the barrel and endcap detectors. Muons are detected in gas-ionization chambers embedded in the steel flux-return yoke outside the solenoid. 
A more detailed description of the CMS detector, together with a definition of the coordinate system and the relevant kinematic variables, 
can be found in Ref.~\cite{CMS:2008xjf}.

Events of interest are selected using a two-tiered trigger system. The first level (L1), composed of custom hardware processors, uses information from 
the calorimeters and muon detectors to select events at a rate of around 100\unit{kHz} within a fixed latency of about 4\mus~\cite{CMS:2020cmk}. 
The second level, known as the high-level trigger (HLT), consists of a farm of processors running a version of the full event reconstruction software 
optimized for fast processing, and reduces the event rate to around 1\unit{kHz} before data storage~\cite{CMS:2016ngn}.

During the 2016 and 2017 data taking, a gradual shift in the timing of the inputs of the ECAL L1 trigger in the region at $\abs{\eta} > 2.0$ 
caused a specific trigger inefficiency~\cite{CMS:2020cmk}. For events containing an electron (a jet) with transverse momentum \pt larger than 
50\GeV (100\GeV), in the region $2.5 < \abs{\eta} < 3.0$ the efficiency loss is $\approx$10--20\%, depending on \pt, $\eta$, and time. 
Correction factors were computed from data and applied to the acceptance evaluated by simulation.

The particle-flow (PF) algorithm~\cite{CMS:2017yfk} reconstructs and identifies each individual particle in an event, 
with an optimized combination of information from the various elements of the CMS detector. The energy of photons is obtained from the ECAL measurement. 
The energy of electrons is determined from a combination of the electron momentum at the primary interaction vertex as determined by the tracker, 
the energy of the corresponding ECAL cluster, and the energy sum of all bremsstrahlung photons spatially compatible with originating from the electron track. 
The energy of muons is obtained from the curvature of the corresponding track. The energy of charged hadrons is determined from a combination of 
their momentum measured in the tracker and the matching ECAL and HCAL energy deposits, corrected for the response function of the calorimeters 
to hadronic showers. Finally, the energy of neutral hadrons is obtained from the corresponding corrected ECAL and HCAL energies. 
The primary vertex (PV) is taken to be the vertex corresponding to the hardest scattering in the event, evaluated using tracking information alone, 
as described in Section 9.4.1 of Ref.~\cite{CMS-TDR-15-02}.

Hadronic jets are clustered from all PF candidates in an event using the infrared and collinear safe anti-\kt 
algorithm~\cite{Cacciari:2008gp, Cacciari:2011ma} with a distance parameter of 0.4. Jet momentum is determined as the vectorial sum of all particle momenta 
in the jet, and, on average, is within 5 to 10\% of the particle level jet momentum over the whole \pt spectrum and detector acceptance. 
Additional $\Pp\Pp$ interactions within the same or nearby bunch crossings (pileup) can contribute additional tracks and calorimetric energy depositions, 
increasing the apparent jet momentum. To mitigate this effect, tracks identified as originating from pileup vertices are discarded and 
a correction is applied to remove the remaining spurious neutral contributions~\cite{Sirunyan:2020foa}.

Jet energy corrections are derived from simulation studies so that the average measured energy of jets becomes identical to that of particle level jets. 
In situ measurements of the momentum balance in dijet, photon$+$jet, $\PZ+$jet, and multijet events are used to determine any residual differences 
between the jet energy scale in data and in simulation, and corrections are applied to simulated events~\cite{CMS:2016lmd}.

The missing transverse momentum vector \ptvecmiss is computed as the negative vector sum of the transverse momenta of all the PF candidates in an event, 
and its magnitude is denoted as \ptmiss~\cite{CMS:2019ctu}. The \ptvecmiss is modified to account for corrections to the energy scale of the 
reconstructed jets in the event. The pileup per particle identification (PUPPI) algorithm~\cite{Bertolini:2014bba}, which weights the PF candidates according 
to their probability to originate from the primary interaction vertex, is applied to reduce the pileup dependence of the \ptvecmiss observable.

\section{Data sets and simulated samples}
\label{sec:MC_samples}

The 2016--2018 data sets, corresponding to integrated luminosities of 36.3, 41.5, and 59.7\fbinv, respectively, are analyzed.
The integrated luminosities for the 2016, 2017, and 2018 data taking years have individual uncertainties of 1.2--2.5\%~\cite{CMS-LUM-17-003,CMS-PAS-LUM-17-004,CMS-PAS-LUM-18-002},
whereas the overall uncertainty for the 2016--2018 period is 1.6\%.
To account for changes in the detector and for the different numbers of pileup interactions, a different 
simulation is employed in the analysis of each yearly data set.

Our analysis requires events filtered by trigger algorithms that select either a single lepton passing a high \pt threshold, 
or two leptons with a lower \pt threshold, satisfying both isolation and identification criteria. 
In the  2016 data set, the \pt threshold of the single electron trigger is  
25\GeV for $\abs{\eta} < 2.1$ and 27\GeV for $2.1 < \abs{\eta} < 2.5$, 
whereas the \pt threshold of the single muon  trigger is 24\GeV. 
Double lepton triggers have lower \pt thresholds,  namely 23\GeV (12\GeV) for the leading (trailing) lepton in the double  electron trigger, 
17\GeV (8\GeV) for the leading (trailing) lepton in the double muon trigger and 23\GeV 
(8\GeV in the first part of the data set, corresponding to 17.7\fbinv, and 12\GeV in the remainder) 
for the  leading (trailing) lepton in the electron-muon trigger. 
In the 2017 data set, single electron and single muon \pt thresholds are raised to  35 and 27\GeV, respectively.
In the 2018 data set the corresponding single lepton \pt thresholds are 32 and 24\GeV. 
Double  lepton \pt thresholds in 2017--2018 data sets are the same as those described for the 2016 data set, 
except for the \pt threshold of the trailing lepton in the electron-muon trigger which is 12\GeV.

All expected physics processes are modeled via Monte Carlo simulation, reweighted to account for known discrepancies between data 
and simulated events. Corrections to the trigger efficiencies, as well as the efficiencies for electron and muon 
reconstruction, identification, and isolation as functions of the lepton \pt and $\eta$, are extracted from events 
with leptonic \PZ boson decays using a ``tag-and-probe'' technique, as described in Ref.~\cite{WZ2011}.
The \PQb jet tagging efficiency~\cite{Bols_2020} is measured and corrections are derived using data samples enriched in \PQb quark 
jets. For each data set, simulated events are reweighted according to the pileup profile distribution observed in data.

The signal process is simulated at leading order (LO) with \MGvATNLO (v.2.4.2)
~\cite{Alwall:2014hca} interfaced with the event generator \PYTHIA 8.240 (8.230)~\cite{Sjostrand:2014zea} in 2016 (2017 and 2018). We require two quarks and two 
leptonically decaying \PW bosons in the final state; contributions from \PGt decays to lighter leptons are also included in the simulation. 
Diagrams containing a top quark contribution are not included in the signal matrix element, since EW top quark
production is taken into account by the \ttbar and single top quark ($\PQt\PW$) background samples.
\PW bosons are generated within 15 decay lifetimes from their on-shell mass. Contributions beyond this limit do not significantly increase the overall cross section 
and therefore are not considered. 

The dipole approach~\cite{sjostrand_2018} is used to model the initial-state radiation, rather than the standard \pt-ordered one used in the \PYTHIA 
parton shower generator, which does not properly describe extra QCD emissions in the vector boson fusion (VBF) and VBS processes~\cite{Jager:2020hkz}.
The QCD-induced \ww background is modeled with \POWHEG v2~\cite{Alioli:2010xd}, and the production of the second jet is described
at LO accuracy in QCD~\cite{Re_2018}. The interference between the EW and QCD-induced processes was evaluated, 
and it results in a negligible effect.

Higgs boson production mechanisms are included in the analysis and treated as a background source. All production modes 
are simulated with \POWHEG v2 at next-to-LO (NLO) accuracy in QCD. Gluon-gluon fusion (ggF) events are further reweighted to match 
next-to-NLO accuracy, according to the NNLOPS scheme~\cite{Hamilton:2015nsa}. The Higgs boson decay into two \PW bosons 
and subsequently into leptons is simulated with the \textsc{JHUgen} generator~\cite{PhysRevD.86.095031}, whereas its decay into 
two \PGt leptons is simulated with \PYTHIA. Other minor Higgs boson decay channels are neglected.

On-shell VBF Higgs boson production is negligible when simulating two on-shell \PW bosons, as is done for the 
signal sample. Accordingly, this contribution is removed from our signal definition, whereas off-shell effects are 
retained. Furthermore, the SR phase space is tailored to enhance EW \ww production occurring without the exchange of a 
Higgs boson, which is suppressed by the tight selection on the dilepton invariant mass as discussed in Section~\ref{sec:Strategy}. 
A dedicated analysis has been designed to target the on-shell VBF Higgs boson production~\cite{CMS-PAS-HIG-20-013}, 
where, conversely, our signal sample is regarded as a background process.

The following additional background processes are modeled in simulation: nonresonant \PW boson pair production induced by gluons 
(included in the QCD-induced \ww background estimation), \ttbar and $\PQt\PW$ production, DY lepton pair production, $\PW\PGg$ and 
$\PZ\PGg$ production, and multiboson production. Most of the event samples are generated at NLO in QCD using either \POWHEG v2, 
\MGvATNLO (v2.4.2), or \MCFM (v7.0)~\cite{Campbell:1999ah, Campbell_2011, campbell2015multithreaded}. 
Only $\PW\PGg$ events are generated in the LO mode with \MGvATNLO (v2.4.2).
The \pt distribution of the \ttbar component of $\ttbar + \PQt\PW$ background is weighted to better match 
data~\cite{Khachatryan:2016mnb}. A similar procedure is applied to the \pt distribution of the \PZ boson in DY samples~\cite{Sirunyan:2019bzr}.
Other EW-induced diboson channels ($\PW\PZ$, $\PW\PGg$ and $\PW^{\pm}\PW^{\pm}$) produced in association with two jets have been evaluated and 
their contribution to the SR is negligible, hence they have not been included in the background estimation. 

The chosen parton distribution functions (PDFs) and underlying event tunes are common to all simulated events for a given data set.
The parton showering and hadronization processes are simulated through \PYTHIA 8.226 (8.230) in 2016 (2017--2018). 
The PDF set is NNPDF 3.0~\cite{Ball:2013hta, Ball:2011uy} (3.1~\cite{Ball:2017nwa}) and the underlying event tune is CUETP8M1~\cite{Khachatryan:2015pea} 
(CP5~\cite{Sirunyan:2019dfx}) for the 2016 (2017--2018) samples. 
All generated events are processed through a simulation of the CMS detector based on \GEANTfour~\cite{GEANT4:2002zbu} 
and are reconstructed with the same algorithms as used for data.

\section{Event selection}
\label{sec:Strategy}

The VBS final state is characterized by the presence of two jets from
the incoming partons with a large \detajj and a high \mjj, 
and two OS leptons and two neutrinos from the \PW boson decays.
Candidate events are preselected if they fulfill the following requirements:
\begin{itemize}
\item Two OS leptons  (electrons or muons), with dilepton mass $\mll > 50\GeV$ and transverse momentum 
      $\ptll > 30\GeV$, are selected with the tight selections described in Refs.~\cite{Sirunyan_2021,Sirunyan:2018fpa}. 
      The thresholds for the highest and second-highest \pt leptons are 25 and 13\GeV, respectively.  
      Events with an additional lepton with $\pt > 10\GeV$ are rejected;
\item \ptmiss $>$ 20\GeV;
\item At least two jets with $\pt > 30\GeV$, $\mjj > 300\GeV$, and $\detajj > 2.5$.
\end{itemize}

Requirements on \mll and \ptll variables are added to reduce contributions from on-shell Higgs boson 
production and DY to \tautau background, respectively, without losing signal efficiency.
The kinematic phase space is then divided into a SR and two CRs, which are used to check the agreement between data and simulation and to 
constrain the normalization of the major backgrounds, \ie, \ttbar and DY production. 
Each region is further categorized according to the charged lepton flavor composition:
two electrons ($\Pe\Pe$), two muons ($\PGm\PGm$), or one electron and one muon ($\Pe\PGm$).
The SR is defined by requiring that no \PQb jets, defined with the loose working point of the DeepJet 
algorithm~\cite{Bols_2020}, are present. The transverse mass \mt is defined as 
$\mT = \sqrt{\smash[b]{2 \ptll \ptmiss \left[1 - \cos\Delta\phi(\ptvecll, \ptvecmiss)\right]}}$, 
where $\phi$ is the azimuthal angle in radians; \mt is required to be above 60\GeV in the $\Pe\PGm$ SR.
In the SR for $\Pe\Pe$ and $\PGm\PGm$, the \ptmiss threshold is raised to 
60\GeV and \mll is required to be greater than 120\GeV to reject DY
\PZ boson production. 
In the $\Pe\PGm$ category, only a residual DY contribution (from $\tautau\to\Pe\PGm$ events) remains.

The SR is further split into two regions to optimize the signal significance. The categorization is 
based on the centrality of the dilepton system with respect to the tagging jets, quantified by the so-called Zeppenfeld 
variable~\cite{Rainwater:1996ud} $\Zll = \frac{1}{2}\abs{\Zlone +
\Zltwo}$, where $\Zl = \etalep - \frac{1}{2}(\etajone + \etajtwo)$
with \etalep, \etajone, and \etajtwo being the pseudorapidities of the
lepton and two jets, respectively.
The categories with $\Zll<1$ are enriched with signal and have less background contamination.
Post-fit event yields are shown in Table~\ref{tab:yields}.

\begin{table*}[ht!]
  \centering
  \topcaption{Post-fit process yields and uncertainties in each SR ($\Pe\Pe$ and $\PGm\PGm$ final states combined).  \label{tab:yields}}
    \begin{tabular}{ccccc}
                          Process   &   SR $\Pe\PGm\ \Zll < 1$  &  SR $\Pe\PGm \Zll > 1$ & SR $\Pe\Pe$ - $\PGm\PGm\ \Zll < 1$ & SR $\Pe\Pe$ - $\PGm\PGm\ \Zll > 1$ \\
      \hline
                             DATA   &               2441    &               2192  &                    1606   &                    1667   \\
              Signal + background   &  $2396.8 \pm 98.5$    & $2239.6 \pm 106.0$  &       $1590.4 \pm 49.4$   &        $1660.5 \pm 43.6$  \\
                            Signal  &   $169.1 \pm 20.2$    &     $69.9 \pm 8.4$  &          $98.0 \pm 6.5$   &           $38.3 \pm 2.5$  \\
                        Background  & $2227.7 \pm 96.4$     &  $2169.7 \pm 105.6$ &      $1492.4 \pm 48.9$    &        $1622.1 \pm 43.5$  \\
                $\ttbar + \PQt\PW$  &  $1629.4 \pm 71.4$    &  $1452.5 \pm 69.5$  &        $767.8 \pm 14.5$   &         $642.5 \pm 13.2$  \\
                    $\PW\PW$ (QCD)  &   $327.0 \pm 61.6$    &   $409.3 \pm 77.3$  &        $111.1 \pm 16.6$   &         $121.5 \pm 17.3$  \\
                        Nonprompt   &   $107.0 \pm 18.4$    &   $109.9 \pm 16.4$  &          $30.0 \pm 4.9$   &           $32.0 \pm 4.2$  \\
                    DY no PU jets   &           \NA         &           \NA       &        $259.5 \pm 27.3$   &         $408.3 \pm 17.1$  \\
                   DY + 1 PU jets   &           \NA         &           \NA       &        $222.7 \pm 33.3$   &         $337.4 \pm 32.9$  \\
                      DY\tautau     &     $69.2 \pm 4.6$    &    $102.0 \pm 5.8$  &                \NA        &                 \NA       \\
                       Multiboson   &     $67.7 \pm 6.6$    &     $75.6 \pm 7.3$  &          $60.9 \pm 3.8$   &           $60.1 \pm 4.8$  \\
                          \PZ{}jj   &      $1.0 \pm 0.2$    &      $0.4 \pm 0.0$  &          $40.5 \pm 4.2$   &           $20.3 \pm 1.3$  \\
                            Higgs   &     $26.6 \pm 1.5$    &     $20.1 \pm 1.0$  &                \NA        &                 \NA       \\
    \end{tabular}
\end{table*}

The \ttbar CRs are defined by inverting the \PQb jet veto and thus requiring the presence of at least one 
\PQb jet with $\pt >20\GeV$ in the final state, resulting in a $\approx95\%$ pure \ttbar sample.
In the DY CRs, the \PQb veto requirement is the same as that in the SR.
In the DY $\Pe\PGm$ category, the $\mt$ requirement is reversed with 
respect to the SR and a $50< \mll <80\GeV$ window is selected. In DY $\Pe\Pe$ and $\PGm\PGm$ categories,
the dilepton mass is chosen to be close to the \PZ boson mass peak, $\abs{\mll - m_\PZ} < 15\GeV$. 
Moreover, the DY $\Pe\Pe$ and $\PGm\PGm$ CRs are divided in two \detajj bins, as explained in 
Section~\ref{sec:SigBkg}.
The fraction of DY events in the DY CRs are $\approx64\%$ and $\approx91\%$ for the $\Pe\PGm$ and $\Pe\Pe$/$\PGm\PGm$ 
categories, respectively. A summary of all regions included in the analysis is shown in \suppMaterial. 

\section{Background estimation and signal extraction}
\label{sec:SigBkg} 
The normalizations of the major backgrounds are measured by a simultaneous fit of data, including CRs. 
For the \ttbar background the determination mainly comes from the dedicated \ttbar CR.

In the $\Pe\Pe$ and $\PGm\PGm$ categories, DY production is one of the leading background sources, 
typically arising when a lepton pair is reconstructed with high \ptmiss due to instrumental effects.
A large fraction of the DY background ($\approx$50\%) comes from events where at least one of the two jets 
originates from a pileup vertex, whereas the remaining DY events are fully associated with the ``hard" interaction, 
in which the two highest \pt jets are radiated by initial-state quarks. These two backgrounds are measured as two 
independent processes and scaled with different parameters during the fit. 
Hence, the normalization of the DY background is determined from separate CRs with $\detajj < 5$ 
(dominated by events where the jets originate from initial-state QCD radiation) and $\detajj > 5$ (dominated by 
events where at least one of the jets comes from a pileup interaction).
A third, minor source of DY background is $\PGt\PGt$ events, and its
normalization is determined from the 
$\Pe\PGm$ category.

In contrast, there is no CR purely enriched by nonresonant QCD-induced \ww events.
The normalization of this background is left to float freely and is
determined by the global fit in all regions.

Nonprompt leptons, \ie,\,either leptons produced in decays of hadrons or jets misidentified as leptons, 
come mainly from \wj events. This background is directly estimated from data by applying a transfer function 
to events entering a \wj CR, where one of the two leptons fails either the tight identification or 
isolation criteria applied in the SR, but passes a looser selection~\cite{Sirunyan:2018egh}.
The transfer function is determined in a separate control sample by measuring the rate of objects satisfying 
loose lepton requirements that also pass the signal lepton criteria. Triggers requiring either one electron and one jet, 
or a single muon are applied to select this control sample. In both cases, the lepton must be well separated from 
the highest \pt jet, and the contribution from leptons originating from \PW boson decays is suppressed by 
requiring $\ptmiss < 20\GeV$. The remaining contamination of prompt leptons from the EW production of a \PZ boson 
in association with jets is estimated from simulation and removed.

Other minor backgrounds, such as the Higgs boson and multiboson production, are estimated entirely from simulation.

In the $\Pe\PGm$ signal category a feed-forward deep neural network (DNN) is used to separate the VBS signal from the \ttbar and 
QCD-induced \ww backgrounds. For optimization purposes, separate DNN models were built for the subregions with low ($\Zll < 1$) 
and high ($\Zll > 1$) values of the Zeppenfeld variable. The two models share the same architecture and are fit using nine 
discriminating variables, which are chosen from a larger set of observables; these variables are listed in Table~\ref{tab:var}. 
The DNN implementation comprises five fully connected hidden layers, the first two (last three) having 128 (64) nodes each, 
that are trained with the stochastic gradient descent technique of the ``Adam" optimizer tool~\cite{kingma_adam:_2017} to achieve a 
good separation of signal and backgrounds. A binary cross-entropy loss function~\cite{binary_crossentropy,Goodfellow-et-al-2016} 
is minimized in both models. The \mjj and \detajj post-fit distributions are shown as an example in \suppMaterial, and they agree 
with SM predictions within post-fit uncertainties. These distributions are also shown in the $\Pe\Pe$ and $\PGm\PGm$ combined 
top CRs, where the data is in excellent agreement with the simulation. The DNN output is illustrated in the $\Pe\PGm$ top CR too; 
less than 5\% residual shape dependence is observed. Such disagreement is mostly concentrated in background-dominated regions of 
the spectrum, and we have checked that it does not affect the signal extraction.

\begin{table}[htb]
  \centering
  \topcaption{Set of variables used as inputs to the DNN for both \Zll $<$ 1 and  \Zll $>$ 1 models. 
           The order in the table corresponds to the importance of the discriminating variable for the \Zll $<$ 1 model, 
           as obtained through the SHAP method~\cite{shap,Shapley:1953ctg}.}
    \cmsTable{
    \begin{tabular}{ll}
    Variable                    & Description                                                 \\
    \hline
    \vspace*{0.7mm}
    $\mjj$                      & Invariant mass of the two tagging jets pair                 \\ \vspace*{0.7mm}
    $\pt^{\mathrm{j}_{1}}$      & \pt of the highest \pt jet                                  \\ \vspace*{0.7mm}
    $\detajj$                   & Pseudorapidity separation  between the two tagging jets     \\ \vspace*{0.7mm}
    $\pt^{\mathrm{j}_{2}}$      & \pt of the second-highest \pt jet                           \\ \vspace*{0.7mm}
    $\Zltwo$                    & Zeppenfeld variable of the second-highest \pt lepton        \\ \vspace*{0.7mm}
    $\ptll $                    & \pt of the lepton pair                                      \\ \vspace*{0.7mm}
    $\Delta\phi _{\ell\ell} $   & Azimuthal angle between the two leptons                     \\ \vspace*{0.7mm}
    $\Zlone$                    & Zeppenfeld variable of the highest \pt lepton               \\ \vspace*{0.7mm}
    $\mt^{\ell_{1}}$            & Transverse mass of the ($\pt^{\ell_{1}}$, \ptmiss) system   \\ 
    \end{tabular}
  \label{tab:var}
  }
  \end{table}

The signal strength modifier of EW \ww production, $\mu_{\mathrm{EW}} = \sobs/\ssm$, is the parameter of interest 
and is translated to a cross section measurement in two different fiducial volumes; more details are given in Section~\ref{sec:results}. 
The signal extraction procedure is based on a binned maximum likelihood fit of the chosen discriminating variable 
distribution, as specified in the next paragraphs, with signal and background templates, performed simultaneously 
in all categories. CRs are included as single-bin templates where the number of events is fit to data.

In the $\Pe\PGm$ SRs, the binned DNN output is chosen as the discriminating variable. The DNN output spans a range between 
0 and 1 and can be interpreted as the probability of each event to be identified as signal. Therefore, high DNN values are 
signal-enriched, whereas background samples mostly populate the low DNN spectrum. In the $\Pe\Pe$ and $\PGm\PGm$ SRs, 
different discriminating variables are chosen as a function of \mjj and \detajj. For $\mjj > 500\GeV$ and $\detajj > 3.5$, 
where the signal-to-background ratio is the largest, \mjj is used. The remaining phase space is divided into three bins 
for each flavor composition ($\Pe\Pe$ and $\PGm\PGm$) and \Zll category (\Zll $<$ 1 and \Zll $>$ 1), the number of events 
in each region being the discriminating variable. The bins are defined as follows:

\begin{itemize}
  \item $300 < \mjj < 500\GeV$ and $ 2.5 < \detajj < 3.5$;
  \item $\mjj > 500\GeV$ and $ 2.5 < \detajj < 3.5$;
  \item $300 < \mjj < 500\GeV$ and $\detajj > 3.5$.
\end{itemize}

The number of events in each bin of the templates included in the likelihood function is modeled as a Poisson 
random variable, with a mean value that is the sum of the contributions from all processes. 
Systematic uncertainties are represented by individual nuisance parameters with log-normal distributions. 
The uncertainties affect the overall normalizations of signal and background processes, as well as the shapes 
of the distributions of the observables. The nuisance parameters associated to shape uncertainties are given by a 
unit Gaussian distribution. Correlations between systematic uncertainties in different categories 
are included. 

\begin{figure*}[!htb]
  \centering
  \includegraphics[width=0.48\textwidth]{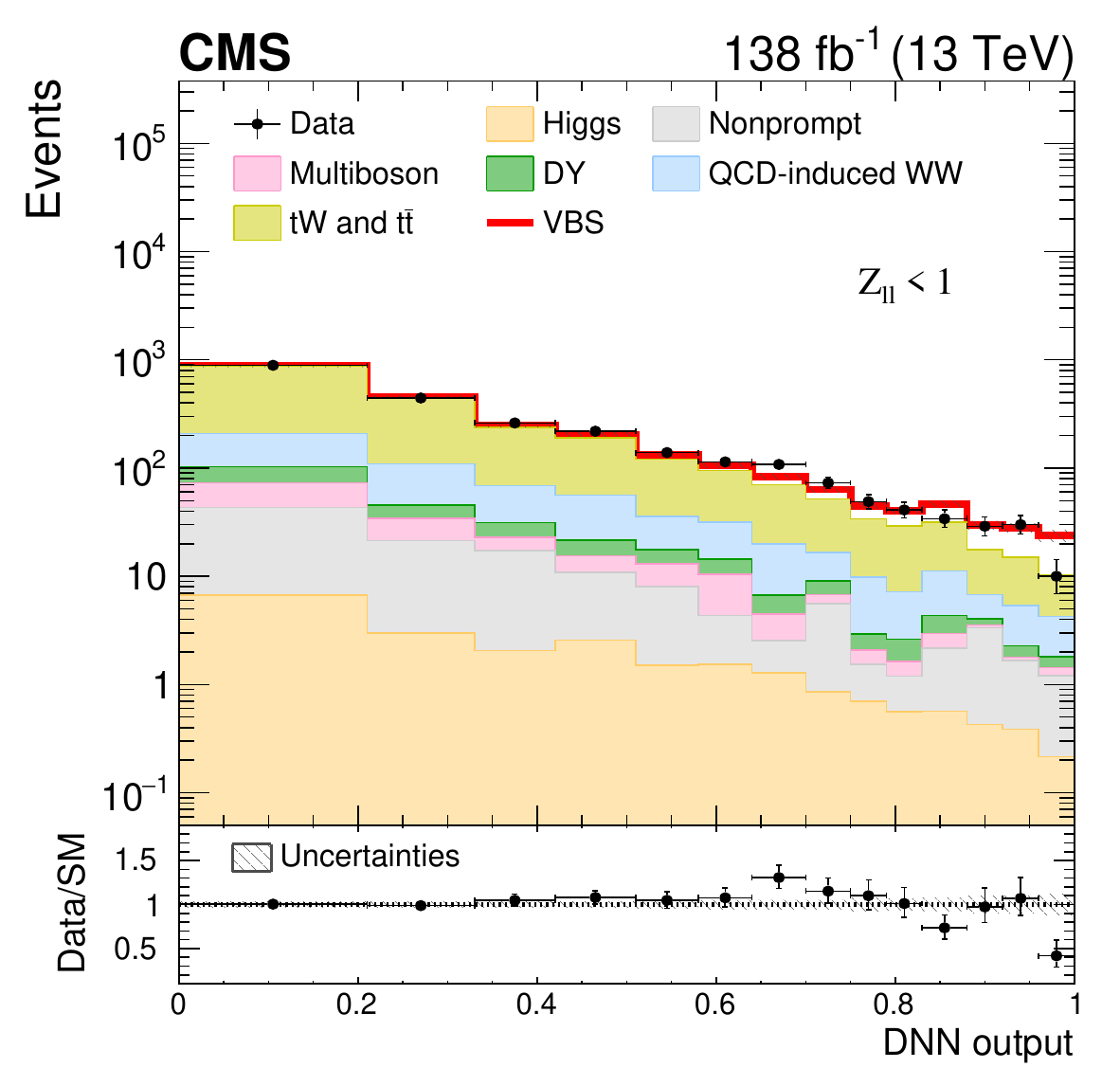}
  \includegraphics[width=0.48\textwidth]{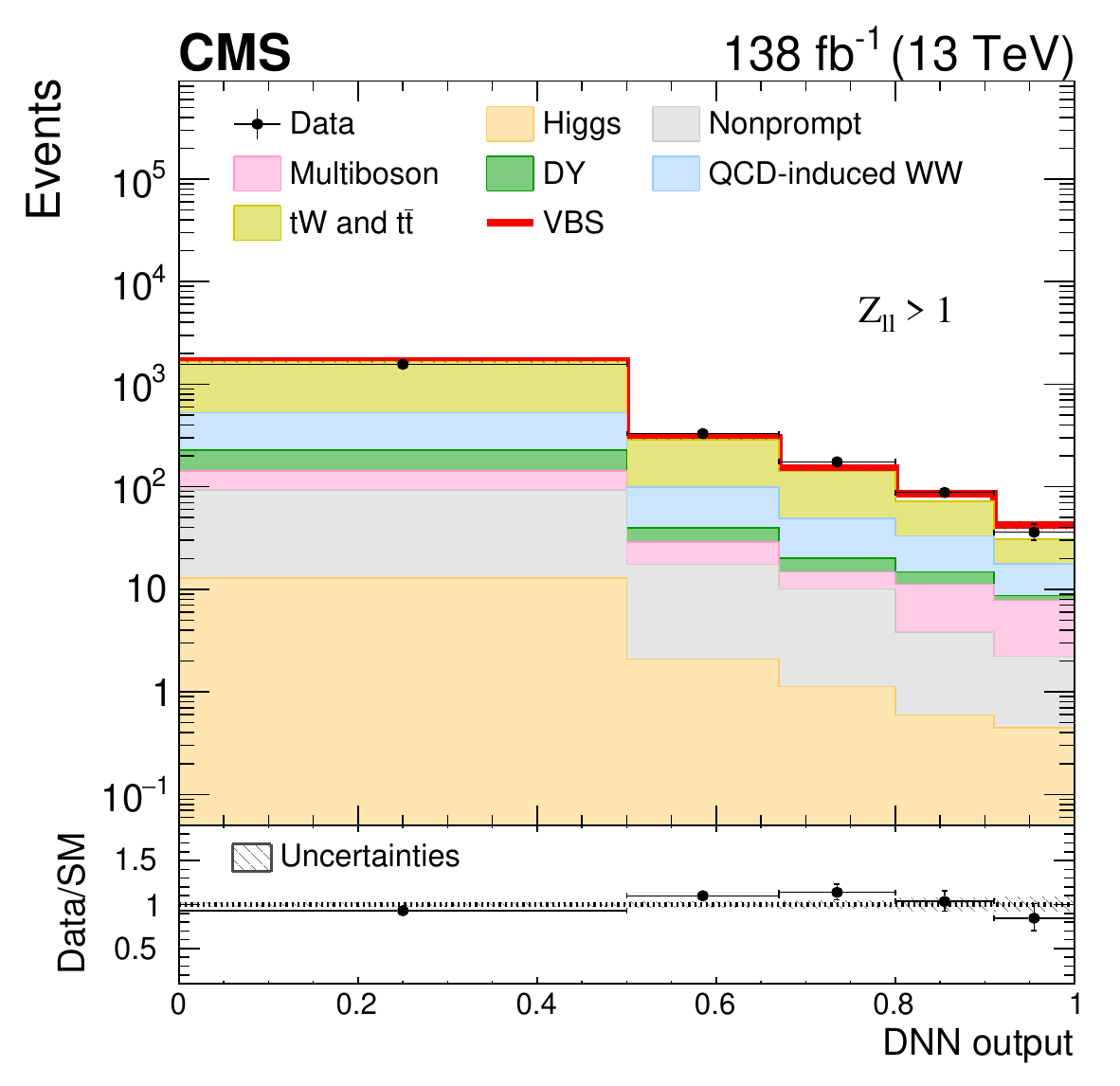}
  \caption{Post-fit DNN output distribution in different-flavor SRs for $\Zll < 1$ (left) and $\Zll > 1$ (right) 
  categories. This variable quantifies how likely each event is signal. The contributions from background and signal 
  (red line) processes are shown as stacked histograms; systematic uncertainties are plotted as dashed gray bands. 
  Data points are displayed with asymmetric Poisson vertical bars to ensure a correct statistical coverage all over the spectrum.}
  \label{fig:VBS_DF}
\end{figure*}

\begin{figure*}[!htb]
  \centering
  \includegraphics[width=0.48\textwidth]{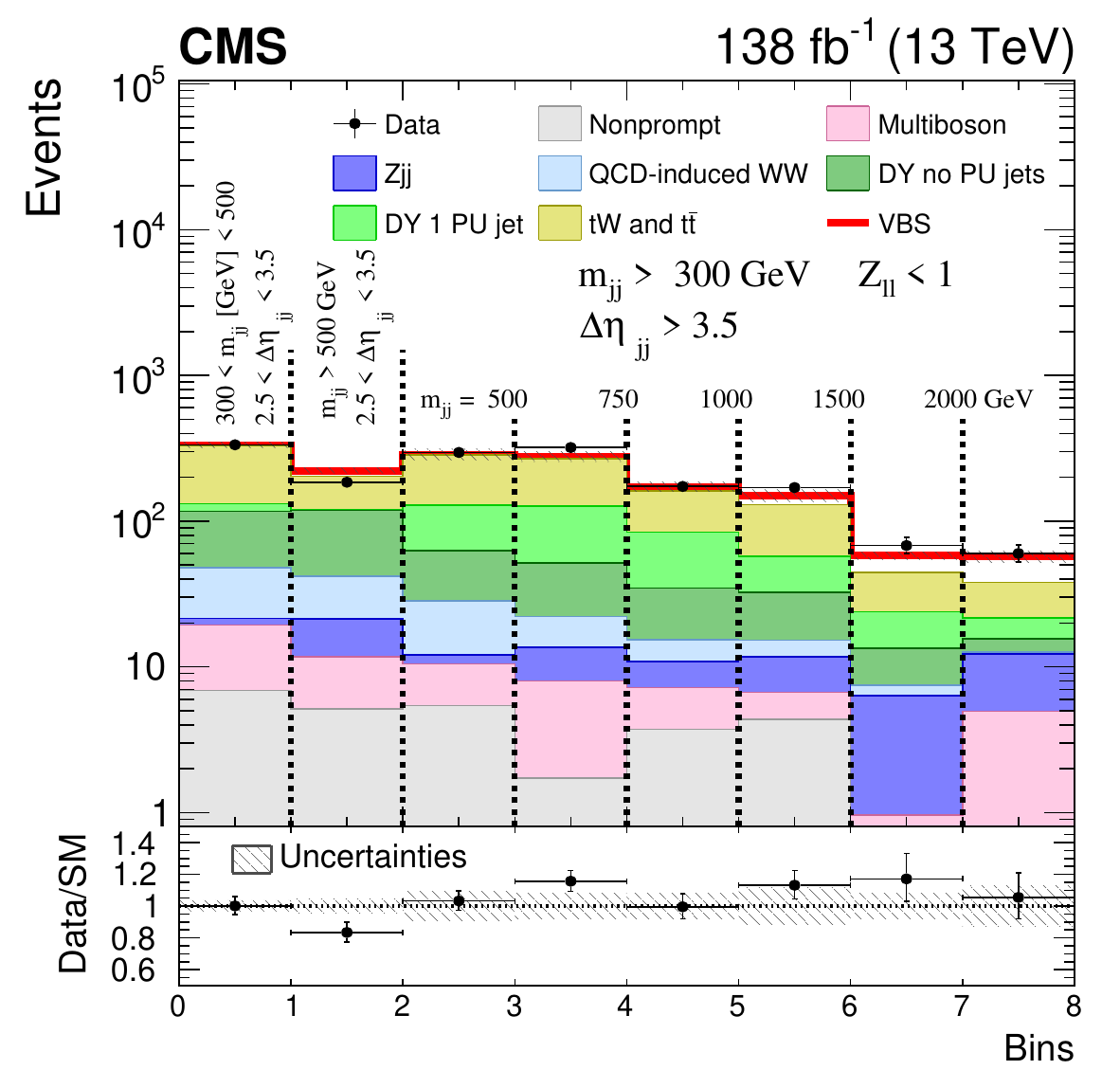}
  \includegraphics[width=0.48\textwidth]{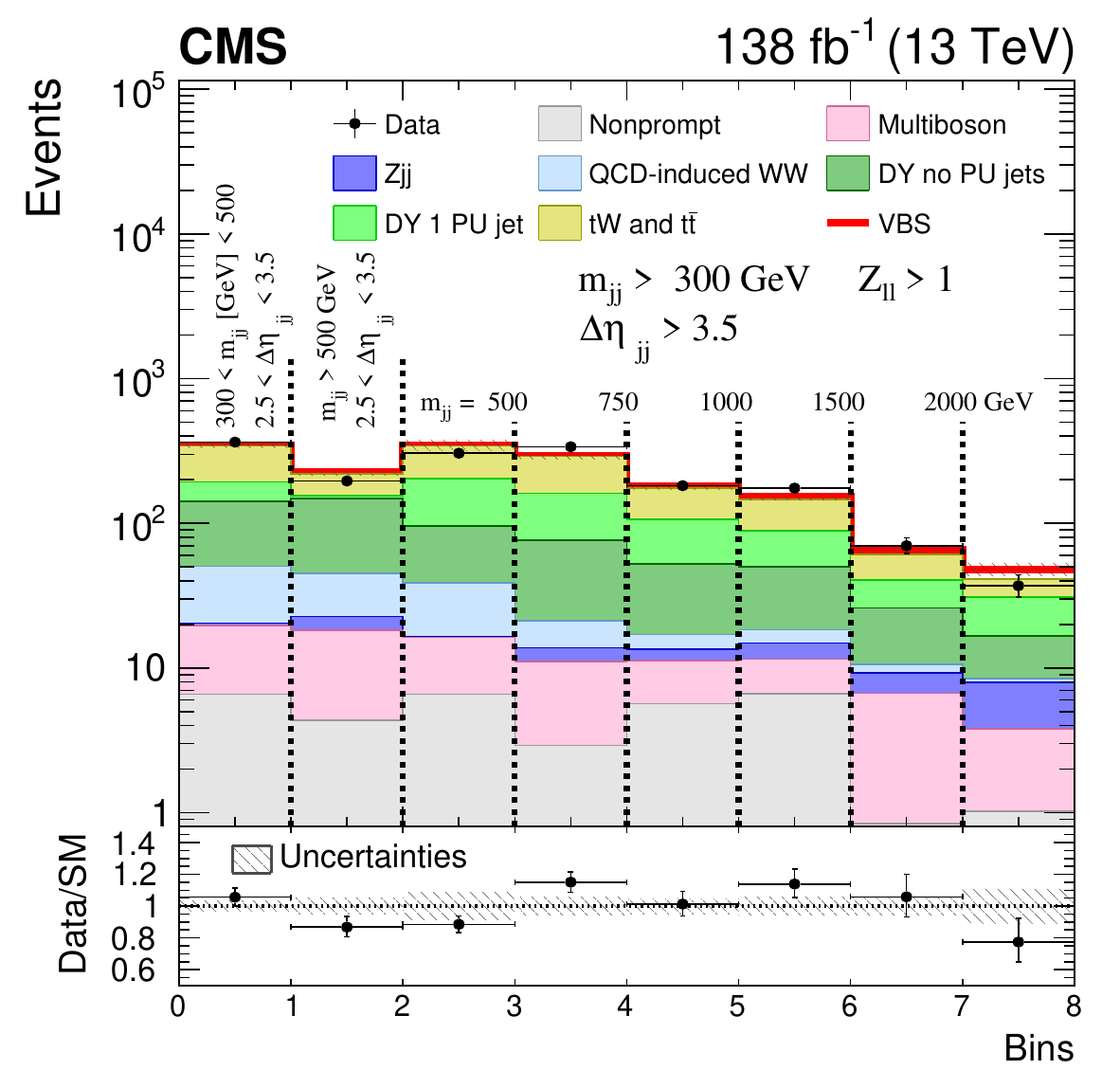}
  \caption{Post-fit \mjj distribution and number of events in same-flavor ($\Pe\Pe$ and $\PGm\PGm$ combined) SRs 
  for $\Zll < 1$ (left) and $\Zll > 1$ (right) categories. The first two bins contain the number of events in 
  the selected region (as reported in the plots themselves). The third bin contains the number of events in the 
  $300 < \mjj[\GeV] < 500$ and $\detajj > 3.5$ regions and, for display purposes, is included in the \mjj 
  distribution, shown in the last five bins. The contributions from background and signal (red line) processes 
  are shown as stacked histograms; systematic uncertainties are plotted as dashed gray bands. Data points are 
  displayed with asymmetric Poisson vertical bars to ensure a correct statistical coverage all over the spectrum.} 
  \label{fig:VBS_SF}
\end{figure*}

\begin{figure*}[!htb]
  \centering
  \includegraphics[width=0.48\textwidth]{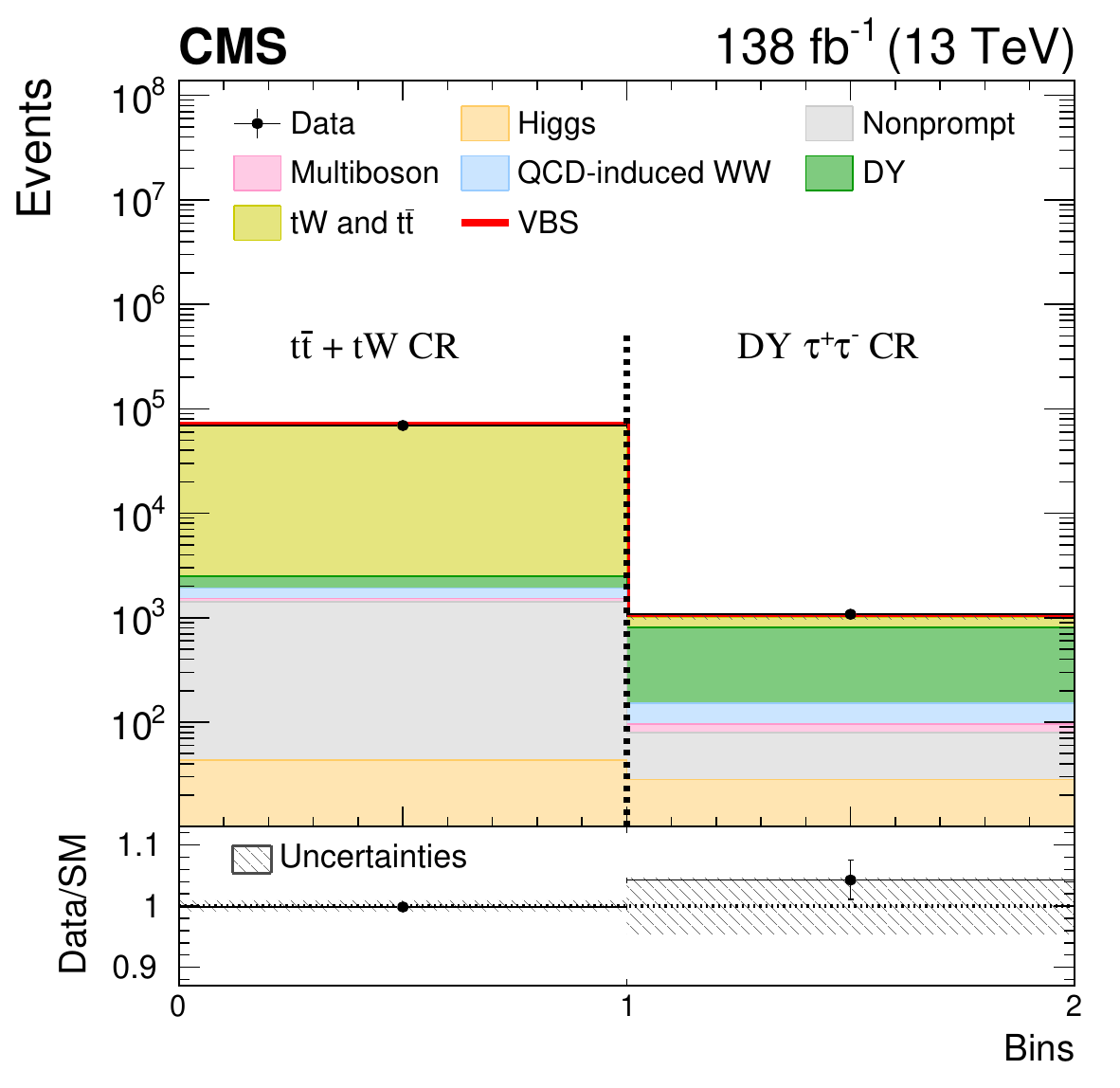}
  \includegraphics[width=0.48\textwidth]{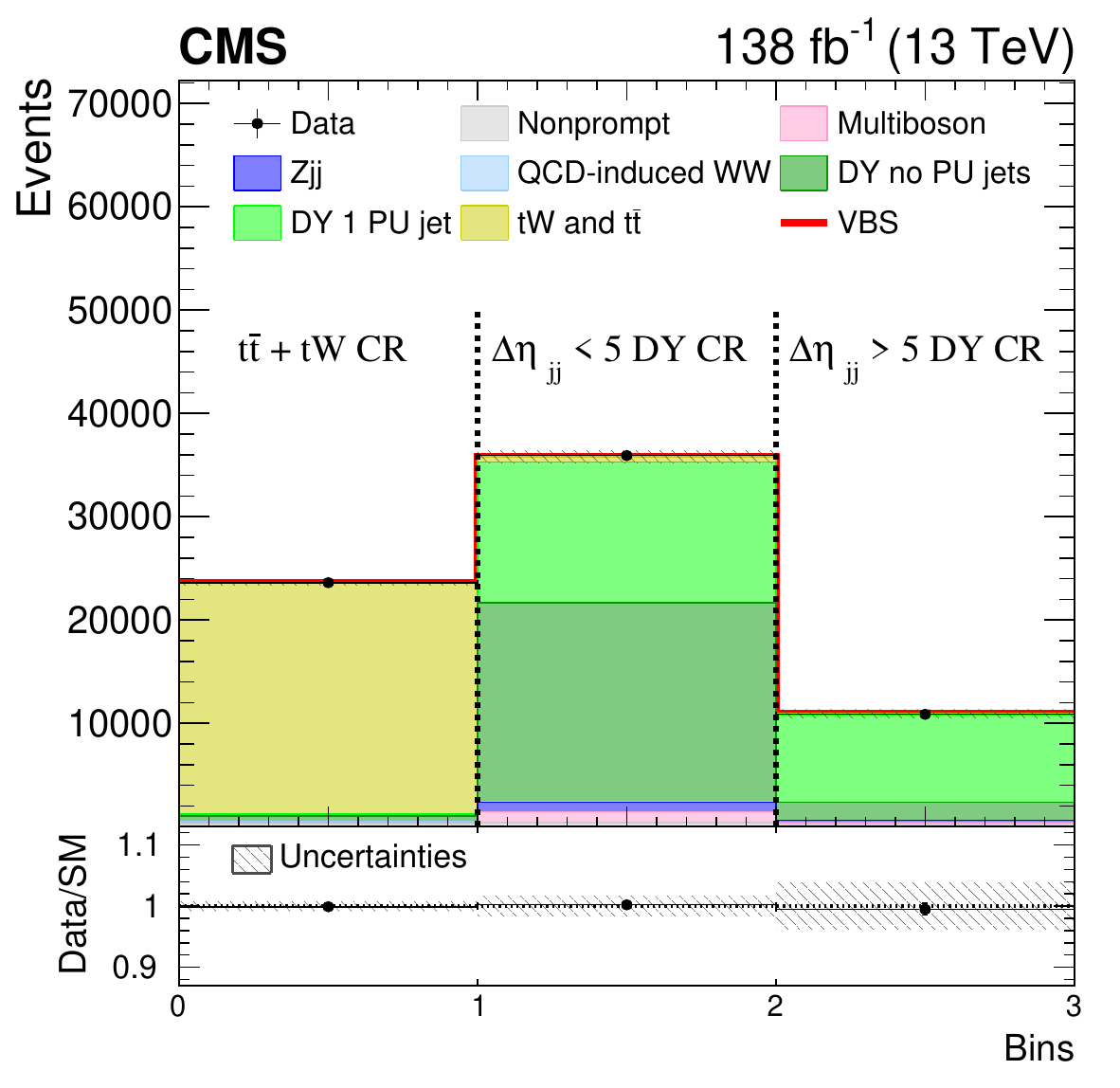}
  \caption{Post-fit number of events in different-flavor (left) and same-flavor (right, with $\Pe\Pe$ and 
  $\PGm\PGm$ combined) CRs. In the left plot, the first bin contains the number of events in the 
  $\ttbar + \PQt\PW$ different-flavor CR, and the second bin those in the DY $\PGt\PGt$ CR. In the right 
  plot, the first bin contains the number of events in the $\ttbar + \PQt\PW$ same-flavor CR, the second bin 
  those in the $\detajj < 5$ DY CR, and the third bin those in the $\detajj > 5$ DY CR. The contributions from 
  background and signal processes (red line) are shown as stacked histograms; systematic uncertainties are plotted as 
  dashed gray bands. Data points are displayed with asymmetric Poisson vertical bars to ensure a correct 
  statistical coverage all over the spectrum.}
  \label{fig:CR}
\end{figure*}

Figs.~\ref{fig:VBS_DF} and~\ref{fig:VBS_SF} show the observed post-fit distributions for the full data set in bins 
of the DNN output for the $\Pe\PGm$ category, and in bins of \mjj and \detajj for the $\Pe\Pe$ and $\PGm\PGm$ categories. 
Similarly, Fig.~\ref{fig:CR} shows the number of post-fit events in the CRs. The difference between data and MC in 
the last bin of the DNN output for \Zll $<$ 1 in Fig.~\ref{fig:VBS_DF} was investigated by verifying that input variables 
reasonably agreed with data at DNN $>$ 0.88. The discrepancy is not localized in any bins of such distributions and, 
because of the good modeling of the top background, is therefore considered to be compatible with a statistical under-fluctuation 
of data.

\section{Systematic uncertainties}
\label{sec:systematics}

Uncertainties in the integrated luminosity, lepton reconstruction and identification efficiency~\cite{CMS:2015xaf,Sirunyan:2018fpa}, 
trigger efficiency and additional trigger timing shift are taken into account. The electron and muon momentum scale uncertainties 
are computed by varying the momenta of leptons within one standard deviation from their nominal value. Similarly, jet energy scale and 
resolution uncertainties~\cite{CMS:2016lmd} are evaluated by shifting the \pt of the jets by one standard deviation, 
and this directly affects the reconstructed jet multiplicity and \ptmiss measurement: several independent sources are considered and 
partially correlated among different data sets. The uncertainties in the residual \ptmiss~\cite{CMS:2019ctu} are also included and 
calculated by varying the momenta of unclustered particles that are not identified with either a jet or a lepton. The \PQb tagging~\cite{Bols_2020} 
introduces various uncertainty sources. The uncertainties from the \PQb tagging algorithm itself are correlated among all data sets, 
whereas uncertainties due to the finite size of the control samples are uncorrelated. Finally, the uncertainty in the pileup reweighting procedure 
is applied to all relevant simulated samples.

Among theoretical uncertainties, effects due to the choice of the renormalization and factorization scales are evaluated, 
to cover missing higher order terms in the perturbation series of cross section calculations. 
These are computed by varying the scales up and down independently by a factor of two with respect to their nominal values, 
ignoring the extreme cases where they are shifted in opposite directions~\cite{Catani:2003zt,Cacciari:2003fi}; 
the envelopes of the various distributions are taken as one standard-deviation variation. Only shape effects are 
included when varying these scales for theoretical uncertainties that affect the signal and main backgrounds, 
since their normalizations are directly measured in data. 

The PDF uncertainties are computed as recommended by the NNPDF prescription~\cite{Rojo:2016ymp}. In the signal process, 
they do not introduce any shape effect in the \mjj and DNN output distributions, hence they are not considered. 
For \ttbar and DY backgrounds, since normalization effects have no impact in the fit, PDF uncertainties can only affect 
the ratio of the expected yields between the SR and the CR. Such uncertainties are included in the CR and estimated to 
be 1\% and 2\% for \ttbar and DY backgrounds, respectively. The modeling of both the parton shower and the underlying 
event is included. For the former, the uncertainties are computed using \PYTHIA by shifting the renormalization 
scale of a factor 2 and 0.5 for both initial and final state radiation; these variations are propagated to each sample, 
but only shape effects are retained for the signal. Underlying event uncertainties are derived 
from alternative samples obtained by varying the \PYTHIA tune parameters as described in Ref.~\cite{Sirunyan:2019dfx}. 
Two fiducial cross section measurements are provided. Therefore, two sets of theoretical uncertainties are computed as 
illustrated above in each fiducial volume.

A list of all systematic uncertainties in the cross section measurement is presented in Table~\ref{tab:syst} and refers 
to the fiducial volume in which signal candidates are selected, as discussed in Section~\ref{sec:results}. 
The systematic component of the overall relative uncertainty in the signal cross section measurement is 13.1\%. 
The statistical uncertainty is evaluated by setting all sources of systematic uncertainty to their best-fit 
result and its value is 14.9\%. The combined relative uncertainty in the cross section measurement is 19.8\%. 

\begin{table}[!htb]
  \centering
  \topcaption{Sources of systematic uncertainty affecting the cross section
    measurement by more than 1\%. The total uncertainty is also reported, 
    as well as the total systematic and statistical contributions.}
  \begin{tabular}{lc}
  Uncertainty source                              & Value       \\
  \hline
  QCD-induced \ww normalization                   & $5.3 \%$    \\
  \ttbar scale variation                          & $5.1 \%$    \\
  VBS signal scale variation                      & $5.0 \%$    \\
  \ttbar normalization                            & $4.9 \%$    \\
  b tagging                                       & $3.5 \%$    \\
  Trigger corrections                             & $3.3 \%$    \\
  DY normalization                                & $2.9 \%$    \\
  Jet energy scale + resolution                   & $2.6 \%$    \\
  Unclustered \ptmiss                             & $2.4 \%$    \\
  QCD-induced \ww scale variation                 & $2.1 \%$    \\
  Integrated luminosity                           & $2.0 \%$    \\
  Muon efficiency                                 & $2.0 \%$    \\
  Pileup                                          & $1.8 \%$    \\
  Electron efficiency                             & $1.5\%$     \\
  Underlying event                                & $1.3 \%$    \\
  Parton shower                                   & $1.0 \%$    \\
  Other                                           & ${<}1\%$     \\
                                                  &             \\
  Total systematic uncertainty                    & $13.1\%$    \\
  Total statistical uncertainty                   & $14.9\%$    \\
  Total uncertainty                               & $19.8\%$    \\
  \end{tabular}
 \label{tab:syst}
\end{table}

\section{Results}
\label{sec:results}

All categories are simultaneously fit to data using a maximum likelihood template fit.
Expected results are assessed by using the Asimov data set~\cite{Cowan_2011}, a pseudoexperiment in which data 
are set in each bin to the value provided by the prediction. 
The statistical significance of the signal is quantified by means of a $p$-value~\cite{Gross_2010}, converted to an equivalent 
Gaussian significance, which corresponds to the probability of observing data with a larger discrepancy with respect to 
the background-only hypothesis, under the asymptotic approximation~\cite{Cowan_2011}. The observed (expected) 
significance for the signal is 5.6\,(5.2) standard deviations.

The EW \ww production cross section is measured in two different fiducial volumes, one more inclusive and 
one closer to the region defined by kinematic criteria applied in the preselection. In the former one, parton-level requirements 
define the phase space of interest, in particular the two outgoing partons ($\PQq\PQqpr$) are required to have $\pt > 10\GeV$ and 
an invariant mass $m_{\PQq\PQqpr} > 100\GeV$; \PGt leptons are included in the simulated signal sample, and their subsequent decay 
to leptons is included as part of the signal. The measured cross section is $99 \pm 20\unit{fb}$, compared with the LO 
prediction of $89 \pm 5\,(\text{scale})\unit{fb}$, where the theoretical uncertainty is computed by varying the factorization scale of the 
signal. 

The other volume is defined as the preselection region where signal candidates are reconstructed, 
as outlined in Section~\ref{sec:Strategy}, but kinematic requirements are transposed to generator-level.
If a photon is found within a distance $\Delta R = \sqrt{\smash[b]{(\Delta\eta)^2+(\Delta\phi)^2}} < 0.1$ from a lepton, its four-momentum is added to that of the 
lepton, making a ``dressed" lepton. Additionally, if such a lepton is found within a distance $\Delta R = 0.4$ from a 
jet axis, the event is discarded. Electrons and muons coming from a \PGt decay are vetoed. The missing transverse momentum is computed 
as the modulus of the vector sum of transverse momenta associated with all invisible particles generated in the event, and is required to be 
greater than $20\GeV$. A summary of the requirements of such a fiducial volume is presented in Table~\ref{tab:fid}. 
The measured fiducial cross section is $10.2 \pm 2.0\unit{fb}$, while the LO theoretical prediction is $9.1 \pm 0.6\,(\text{scale})\unit{fb}$.

\begin{table}[htb]
  \centering
  \topcaption{Definition of the fiducial volume similar to the reconstructed SR.}
  \cmsTable{
    \begin{tabular}{ll}
    Objects                             & Requirements    \\
    \hline
    \vspace{0.7mm}
    \multirow{5}{*}{Leptons}            & $\Pe\PGm, \Pe\Pe, \PGm\PGm$ (not from \PGt decay), opposite charge                                          \\ \vspace{0.7mm}
                                        & $\pt ^{\text{ dressed}\:\ell}= \pt ^{\ell} + \sum_{i}{\pt ^{\gamma_i}}$ if $\Delta R(\ell,\gamma_i) < 0.1$  \\ \vspace{0.7mm}
                                        & $\ptlone > 25\GeV$, $\ptltwo > 13\GeV$, $\pt ^{\ell_3} < 10\GeV$                                            \\ \vspace{0.7mm}
                                        & $\abs{\eta} < 2.5$                                                                                          \\ \vspace{0.7mm}
                                        & $\ptll > 30\GeV$, $\mll > 50\GeV$                                                                           \\ \vspace{0.7mm}
                                        &                                                                                                             \\ \vspace{0.7mm}
    \multirow{5}{*}{Jets}               & $\pt ^j > 30\GeV$                                                                                           \\ \vspace{0.7mm}
                                        & $\Delta R (j, \ell) > 0.4 $                                                                                 \\ \vspace{0.7mm}
                                        & At least 2 jets, no \PQb jets                                                                               \\ \vspace{0.7mm}
                                        & $\abs{\eta} < 4.7$                                                                                          \\ \vspace{0.7mm}
                                        & $\mjj > 300\GeV$, $\detajj > 2.5$                                                                           \\ \vspace{0.7mm}
                                        &                                                                                                             \\ \vspace{0.7mm}
    $\ptmiss$                           & $\ptmiss > 20\GeV$                                                                                          \\ \vspace{0.7mm}
    \end{tabular}
  \label{tab:fid}
  }
  \end{table}

\section{Summary}
\label{sec:summary}

The EW production of a pair of opposite-sign \PW bosons in association with two jets has been observed 
in a data set corresponding to an integrated luminosity of 138\fbinv collected with the CMS detector 
at the CERN LHC in proton-proton collisions at $\sqrts = 13\TeV$. Tabulated results are provided 
in the HEPData record for this analysis~\cite{hepdata}. Events containing two opposite-sign leptons (electrons or muons), 
missing transverse momentum, and two jets with large separation in pseudorapidity and high dijet
invariant mass were selected. A deep neural network was employed to deal with the irreducible background 
from the QCD-induced production of \PW boson pairs, and the dominant background from the production of \ttbar quark pairs.

The measured signal corresponds to an observed (expected) significance of 5.6\,(5.2) standard deviations with respect to 
the background-only hypothesis. The EW \ww production cross section has been measured in two fiducial volumes. 
In the more inclusive one, the cross section is $99 \pm 20\unit{fb}$ ($89 \pm 5\unit{fb}$ expected), 
whereas in that comparable with the experimental phase space the measured cross section is $10.2 \pm 2.0\unit{fb}$ 
($9.1 \pm 0.6\unit{fb}$ expected). These results are compatible with standard model predictions within one standard deviation.

\begin{acknowledgments}

We congratulate our colleagues in the CERN accelerator departments for the excellent performance of the LHC and thank the technical and administrative staffs at CERN and at other CMS institutes for their contributions to the success of the CMS effort. In addition, we gratefully acknowledge the computing centers and personnel of the Worldwide LHC Computing Grid and other centers for delivering so effectively the computing infrastructure essential to our analyses. Finally, we acknowledge the enduring support for the construction and operation of the LHC, the CMS detector, and the supporting computing infrastructure provided by the following funding agencies: BMBWF and FWF (Austria); FNRS and FWO (Belgium); CNPq, CAPES, FAPERJ, FAPERGS, and FAPESP (Brazil); MES and BNSF (Bulgaria); CERN; CAS, MoST, and NSFC (China); MINCIENCIAS (Colombia); MSES and CSF (Croatia); RIF (Cyprus); SENESCYT (Ecuador); MoER, ERC PUT and ERDF (Estonia); Academy of Finland, MEC, and HIP (Finland); CEA and CNRS/IN2P3 (France); BMBF, DFG, and HGF (Germany); GSRI (Greece); NKFIH (Hungary); DAE and DST (India); IPM (Iran); SFI (Ireland); INFN (Italy); MSIP and NRF (Republic of Korea); MES (Latvia); LAS (Lithuania); MOE and UM (Malaysia); BUAP, CINVESTAV, CONACYT, LNS, SEP, and UASLP-FAI (Mexico); MOS (Montenegro); MBIE (New Zealand); PAEC (Pakistan); MES and NSC (Poland); FCT (Portugal); MESTD (Serbia); MCIN/AEI and PCTI (Spain); MOSTR (Sri Lanka); Swiss Funding Agencies (Switzerland); MST (Taipei); MHESI and NSTDA (Thailand); TUBITAK and TENMAK (Turkey); NASU (Ukraine); STFC (United Kingdom); DOE and NSF (USA).
If acknowledgements for individuals are required for a short letter because some of the principal authors are funded through individual grants, it should be OK to add the lines below concerning individuals even for a short letter, but please first consult with the PubComm chair.

\hyphenation{Rachada-pisek} Individuals have received support from the Marie-Curie program and the European Research Council and Horizon 2020 Grant, contract Nos.\ 675440, 724704, 752730, 758316, 765710, 824093, 884104, and COST Action CA16108 (European Union); the Leventis Foundation; the Alfred P.\ Sloan Foundation; the Alexander von Humboldt Foundation; the Belgian Federal Science Policy Office; the Fonds pour la Formation \`a la Recherche dans l'Industrie et dans l'Agriculture (FRIA-Belgium); the Agentschap voor Innovatie door Wetenschap en Technologie (IWT-Belgium); the F.R.S.-FNRS and FWO (Belgium) under the ``Excellence of Science -- EOS" -- be.h project n.\ 30820817; the Beijing Municipal Science \& Technology Commission, No. Z191100007219010; the Ministry of Education, Youth and Sports (MEYS) of the Czech Republic; the Hellenic Foundation for Research and Innovation (HFRI), Project Number 2288 (Greece); the Deutsche Forschungsgemeinschaft (DFG), under Germany's Excellence Strategy -- EXC 2121 ``Quantum Universe" -- 390833306, and under project number 400140256 - GRK2497; the Hungarian Academy of Sciences, the New National Excellence Program - \'UNKP, the NKFIH research grants K 124845, K 124850, K 128713, K 128786, K 129058, K 131991, K 133046, K 138136, K 143460, K 143477, 2020-2.2.1-ED-2021-00181, and TKP2021-NKTA-64 (Hungary); the Council of Science and Industrial Research, India; the Latvian Council of Science; the Ministry of Education and Science, project no. 2022/WK/14, and the National Science Center, contracts Opus 2021/41/B/ST2/01369 and 2021/43/B/ST2/01552 (Poland); the Funda\c{c}\~ao para a Ci\^encia e a Tecnologia, grant CEECIND/01334/2018 (Portugal); the National Priorities Research Program by Qatar National Research Fund; MCIN/AEI/10.13039/501100011033, ERDF ``a way of making Europe", and the Programa Estatal de Fomento de la Investigaci{\'o}n Cient{\'i}fica y T{\'e}cnica de Excelencia Mar\'{\i}a de Maeztu, grant MDM-2017-0765 and Programa Severo Ochoa del Principado de Asturias (Spain); the Chulalongkorn Academic into Its 2nd Century Project Advancement Project, and the National Science, Research and Innovation Fund via the Program Management Unit for Human Resources \& Institutional Development, Research and Innovation, grant B05F650021 (Thailand); the Kavli Foundation; the Nvidia Corporation; the SuperMicro Corporation; the Welch Foundation, contract C-1845; and the Weston Havens Foundation (USA).
\end{acknowledgments}

\bibliography{auto_generated}

\ifthenelse{\boolean{cms@external}}{}{
\clearpage
\appendix
\section{Supplementary material for SMP-21-001\label{app:suppMat}}
\input{supplemental_material}
}

\cleardoublepage \section{The CMS Collaboration \label{app:collab}}\begin{sloppypar}\hyphenpenalty=5000\widowpenalty=500\clubpenalty=5000\input{SMP-21-001-public-authorlist.tex}\end{sloppypar}
\end{document}

%% file: supplemental_material.tex
\begin{table*}[htb!]
  \centering
  \topcaption{Summary of the event categorization on top of signal candidates preselection. In each region, same-flavor final states share the same kinematic requirements.   \label{tab:aux_categorization}}
    \begin{tabular}{cccc}
      Region                                  & Final state                   & Requirements                                & Subregion                            \\
      \hline
      \multirow{9}{*}{SR}                     & \multirow{2}{*}{$\Pe\PGm/\PGm\Pe$}  & $\mt > 60\GeV$                             & $\Zll < 1$                            \\ 
                                              &                               & $\mll >50\GeV$                             & $\Zll > 1$                            \\
                                              &                               & no \PQb jet with $\pt > 20\GeV$            &                                       \\
                                              &                               &                                            &                                       \\
                                              & \multirow{2}{*}{$\Pe\Pe$}           &                                            & $\Zll < 1$                            \\ 
                                              &                               & $\mll >120\GeV$                            & $\Zll > 1$                            \\
                                              &                               & $\ptmiss > 60\GeV$                         &                                       \\
                                              & \multirow{2}{*}{$\PGm\PGm$}     & no \PQb jet with $\pt > 20\GeV$            & $\Zll < 1$                            \\ 
                                              &                               &                                            & $\Zll > 1$                            \\ \hline
      \multirow{6}{*}{$\ttbar + \PQt\PW$  CR} & \multirow{2}{*}{$\Pe\PGm/\PGm\Pe$}  & $\mll >50\GeV$                             &                                       \\
                                              &                               & at least one \PQb jet with $\pt > 20\GeV$  &                                       \\
                                              &                               &                                            &                                       \\
                                              & $\Pe\Pe$                            & $\mll > 120\GeV$                           &                                       \\
                                              &                               & $\ptmiss > 60\GeV$                         &                                       \\
                                              & $\PGm\PGm$                      & at least one \PQb jet with $\pt > 20\GeV$  &                                       \\ \hline
      \multirow{9}{*}{DY CR}                  & \multirow{3}{*}{$\Pe\PGm/\PGm\Pe$}  & $\mt < 60\GeV$                             &                                       \\
                                              &                               & $50\GeV < \mll < 80\GeV$                   &                                       \\
                                              &                               & no \PQb jet with $\pt > 20\GeV$            &                                       \\
                                              &                               &                                            &                                       \\
                                              & \multirow{2}{*}{$\Pe\Pe$}           &                                            & $\detajj < 5$                         \\
                                              &                               & $\abs{\mll - m_Z} < 15\GeV$                & $\detajj > 5$                         \\ 
                                              &                               & $\ptmiss > 60\GeV$                         &                                       \\
                                              & \multirow{2}{*}{$\PGm\PGm$}     & no \PQb jet with $\pt > 20\GeV$            & $\detajj < 5$                         \\ 
                                              &                               &                                            & $\detajj > 5$                         \\
    \end{tabular}
\end{table*}

\begin{figure*}[htbp!]
  \centering
  \includegraphics[width=0.48\textwidth]{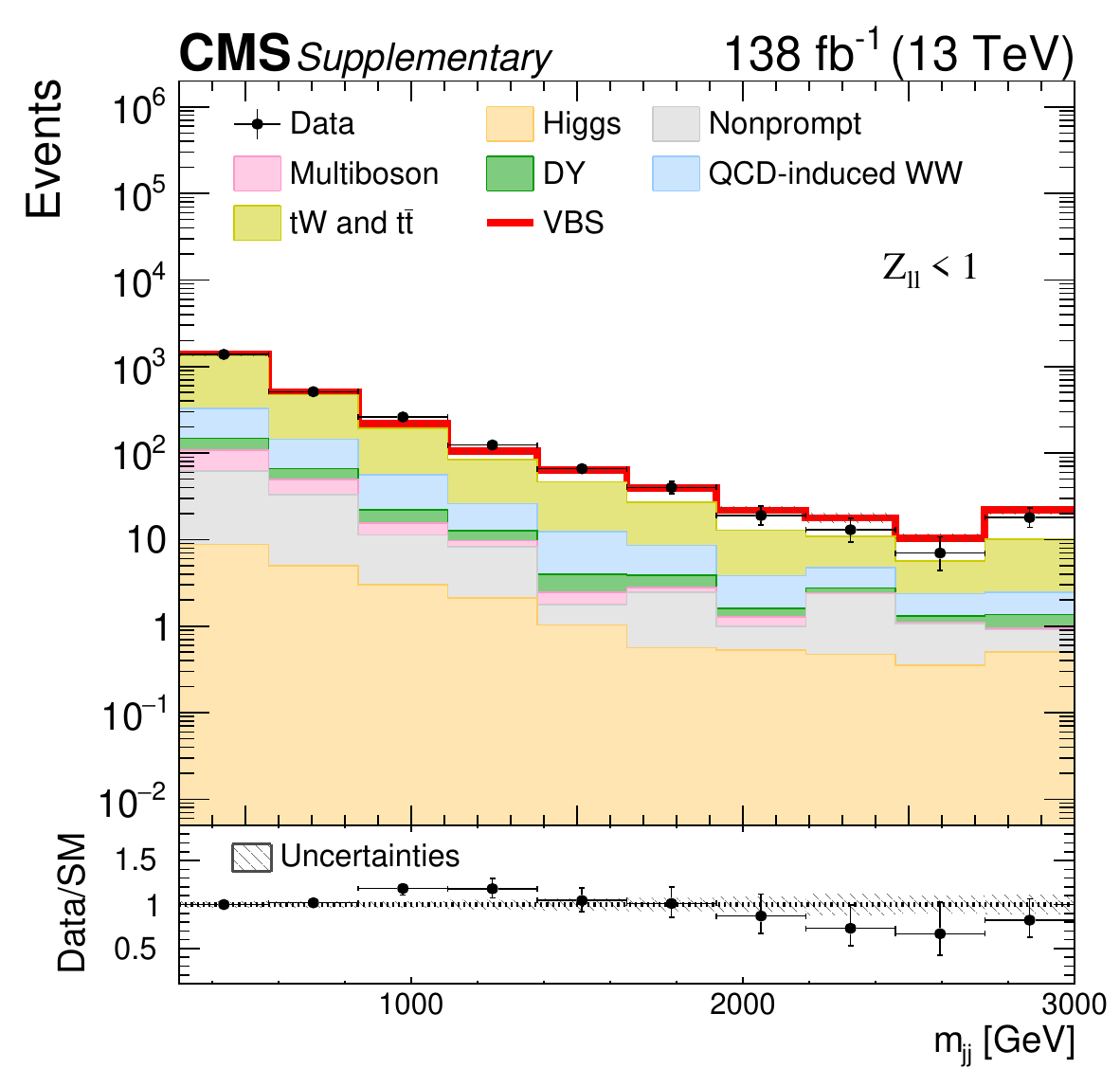}
  \includegraphics[width=0.48\textwidth]{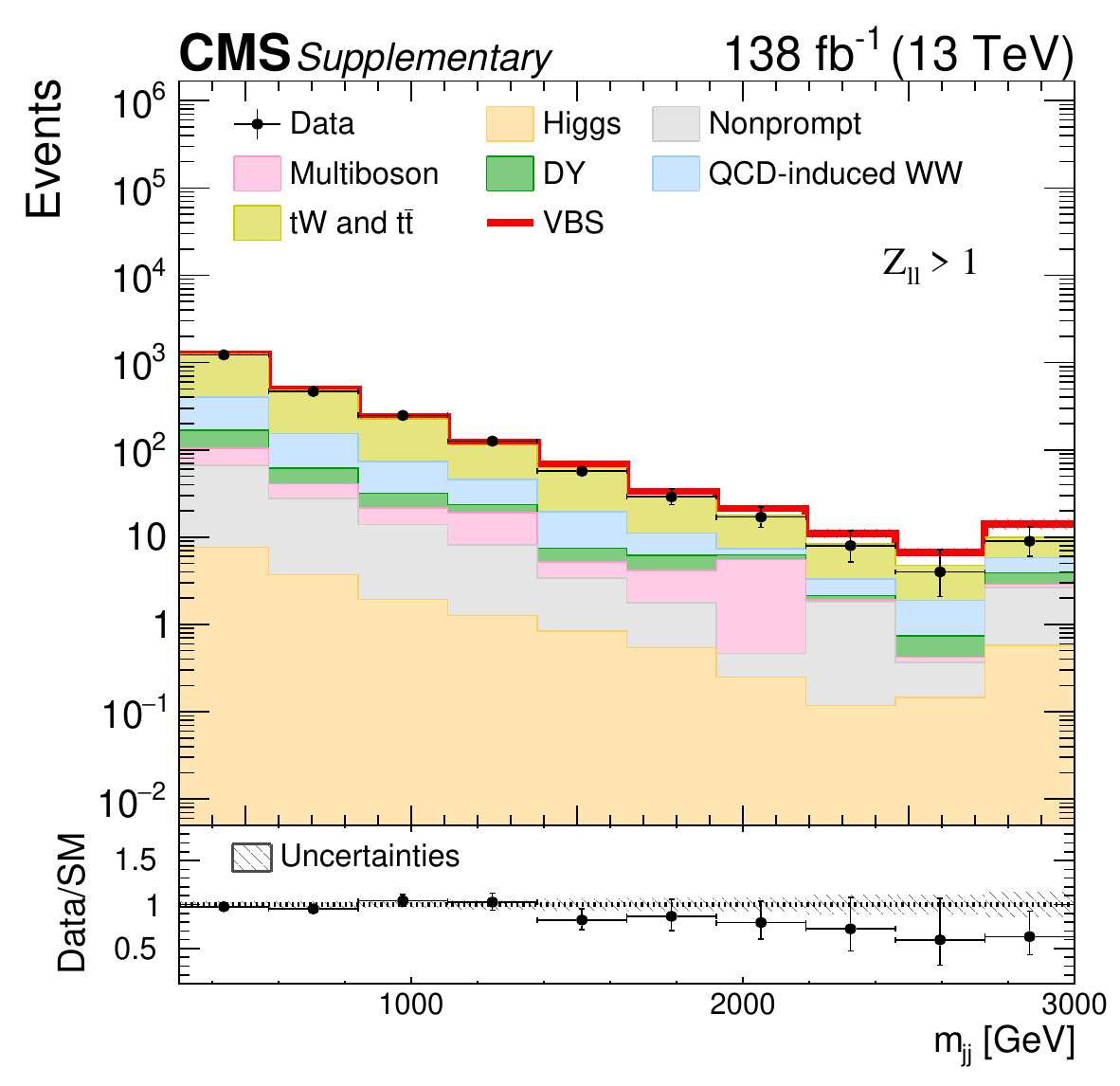}
  \includegraphics[width=0.48\textwidth]{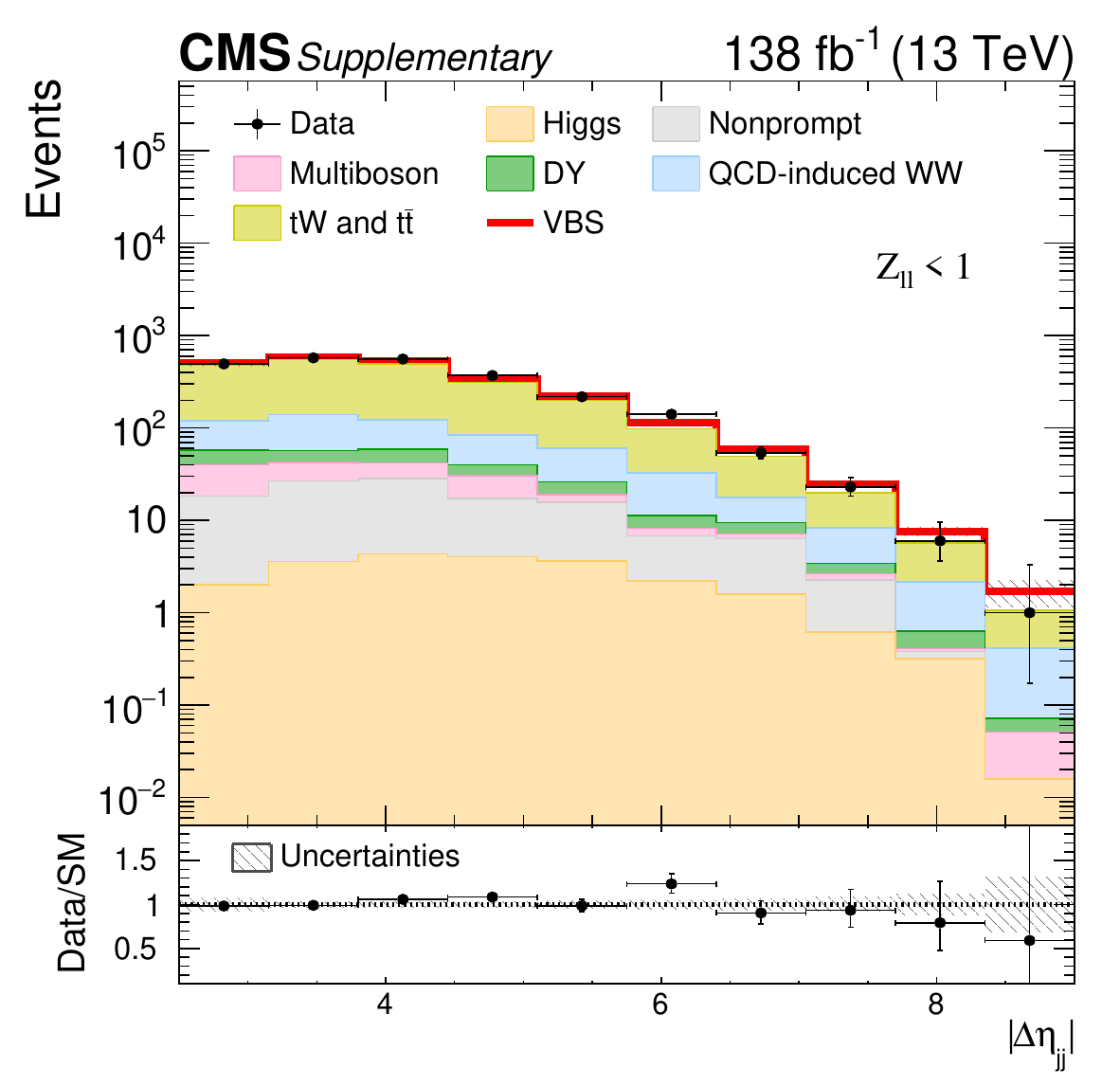}
  \includegraphics[width=0.48\textwidth]{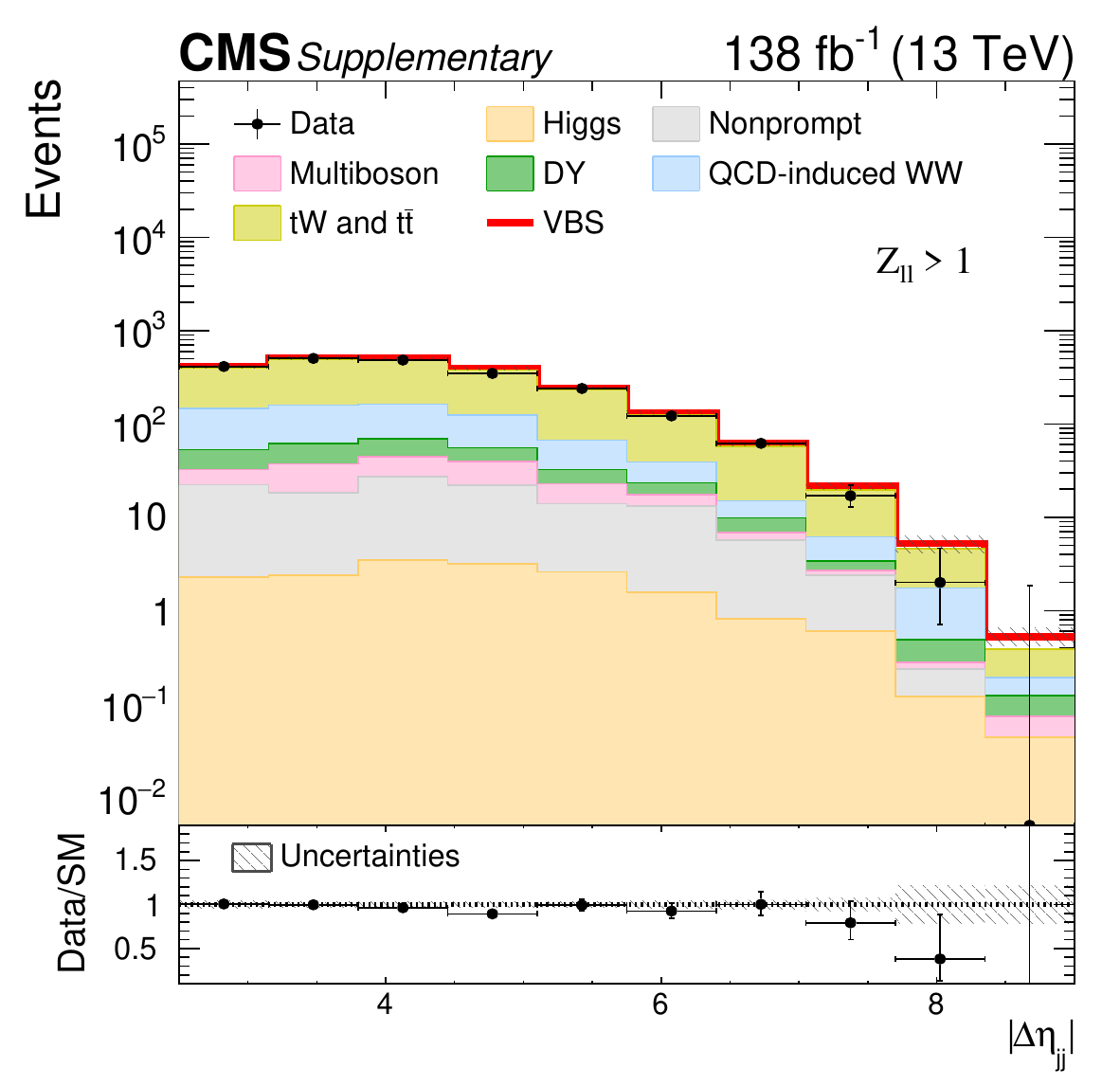}
  \caption{Post-fit distributions of \mjj (upper row) and \detajj (lower row) variables in different-flavor SRs for $\Zll < 1$ (left column) 
  and $\Zll > 1$ (right column) categories. These variables are among the nine observables used as inputs for the DNN, as listed in Table~2.
  The contributions from background and signal (red line) processes are shown as stacked histograms; systematic uncertainties are plotted as dashed gray bands. 
  Data points are displayed with asymmetric Poisson vertical bars to ensure a correct statistical coverage all over the spectrum.  \label{fig:aux_DNN}}
\end{figure*}

\begin{figure*}[htbp!]
  \centering
  \includegraphics[width=0.48\textwidth]{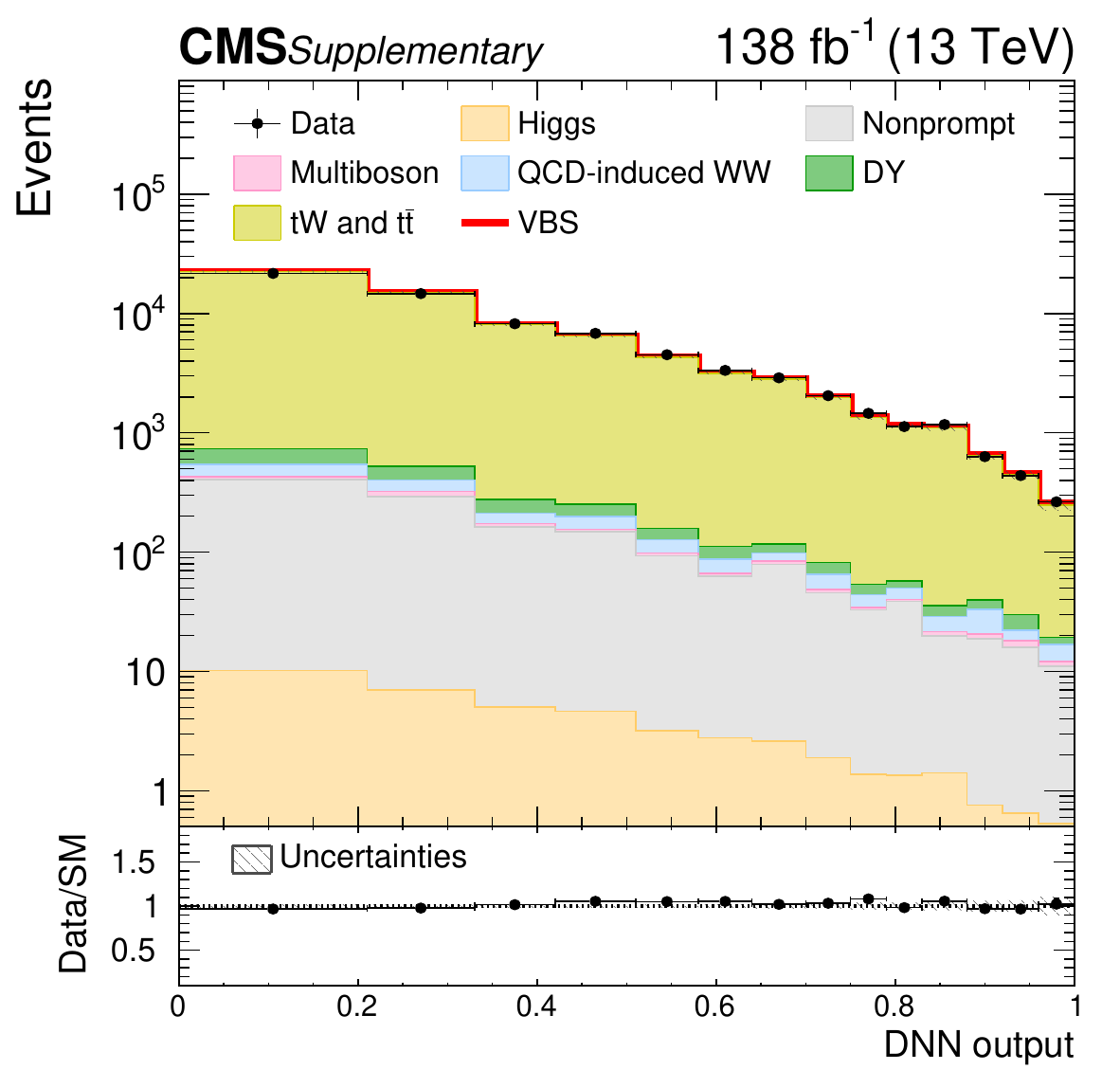}
  \caption{Post-fit distribution of the DNN output in different-flavor top CR. Data to SM expectation ratio has some residual shape dependence 
  (less than 5\% across the full spectrum), which however was found to not affect the analysis. The contributions from background and signal 
  (red line) processes are shown as stacked histograms; systematic uncertainties are plotted as dashed gray bands. Data points are displayed 
  with asymmetric Poisson vertical bars to ensure a correct statistical coverage all over the spectrum.  \label{fig:aux_DNN_topCR}}
\end{figure*}

\begin{figure*}[htbp!]
  \centering
  \includegraphics[width=0.48\textwidth]{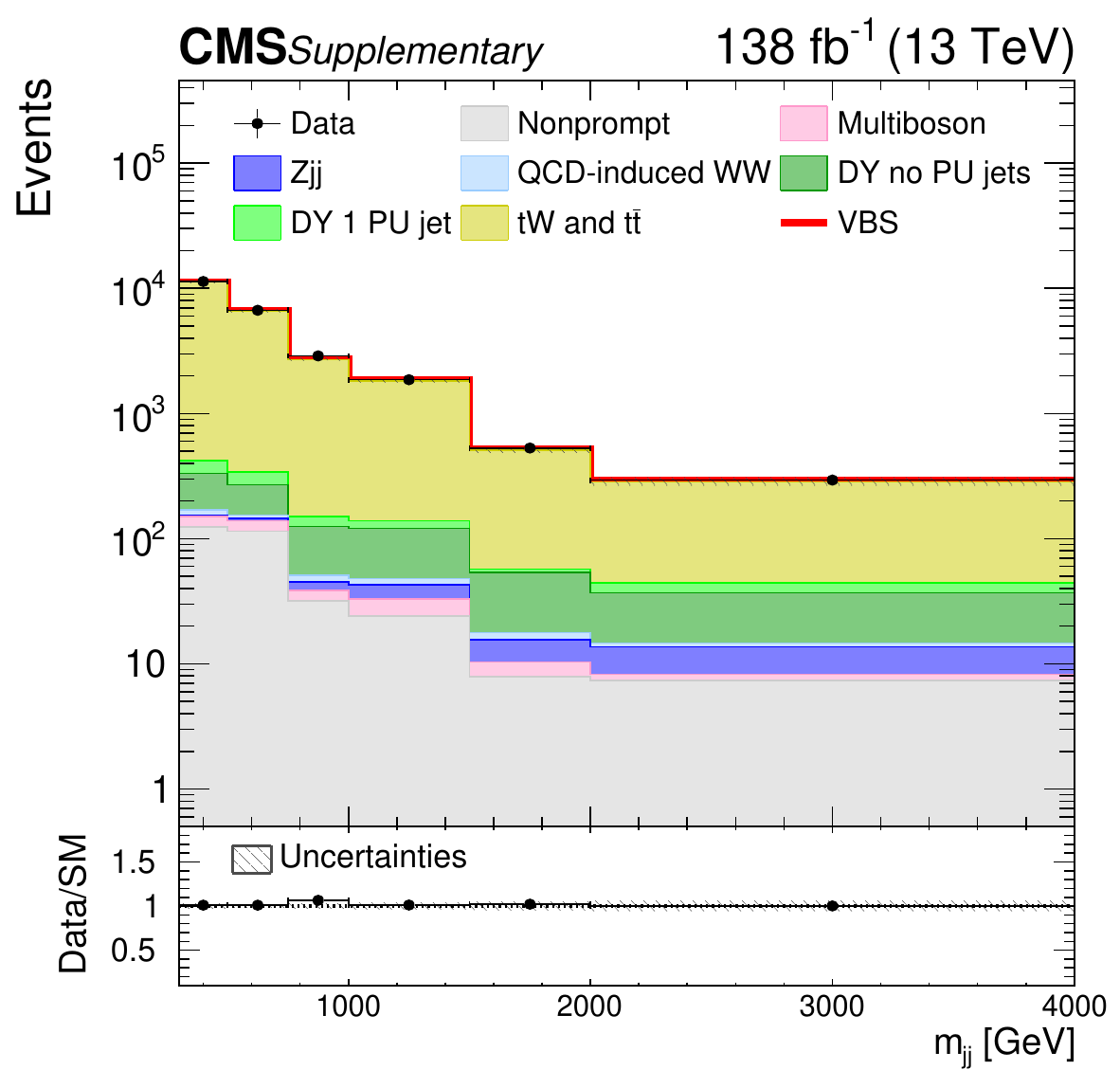}
  \includegraphics[width=0.48\textwidth]{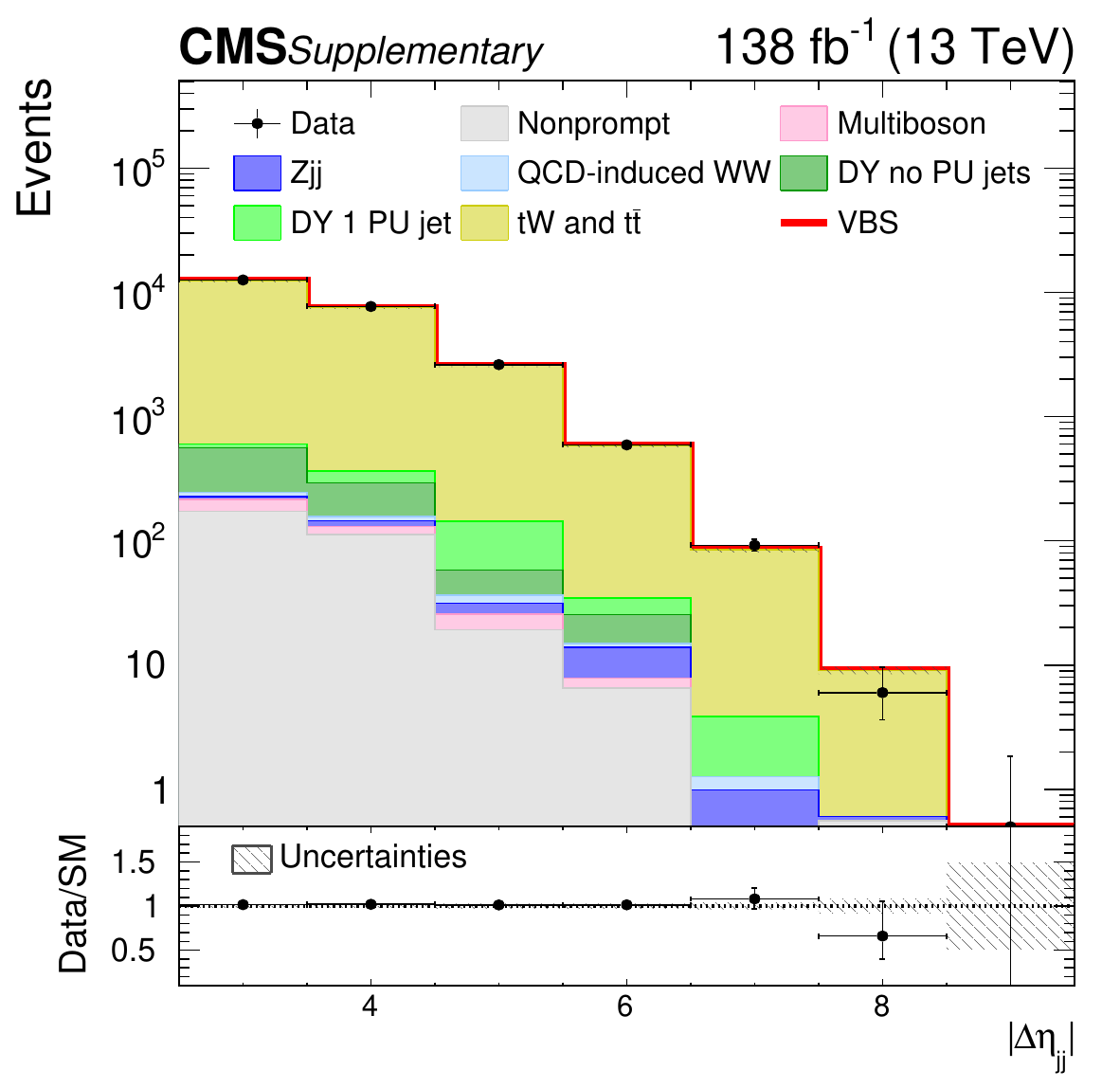}
  \caption{Post-fit distributions of \mjj (left) and \detajj (right) variables in same-flavor top CR ($\Pe\Pe$ and $\PGm\PGm$ final states combined). 
  The ratio between data and SM expectation is in excellent agreement with 1 within post-fit uncertainties in both spectra. 
  The contributions from background and signal (red line) processes are shown as stacked histograms; systematic uncertainties are plotted as dashed gray bands. 
  Data points are displayed with asymmetric Poisson vertical bars to ensure a correct statistical coverage all over the spectrum.  \label{fig:aux_SF_topCR}}
\end{figure*}

%% file: SMP-21-001-public-authorlist.tex
\cmsinstitute{Yerevan Physics Institute, Yerevan, Armenia}
{\tolerance=6000
A.~Tumasyan\cmsAuthorMark{1}\cmsorcid{0009-0000-0684-6742}
\par}
\cmsinstitute{Institut f\"{u}r Hochenergiephysik, Vienna, Austria}
{\tolerance=6000
W.~Adam\cmsorcid{0000-0001-9099-4341}, J.W.~Andrejkovic, T.~Bergauer\cmsorcid{0000-0002-5786-0293}, S.~Chatterjee\cmsorcid{0000-0003-2660-0349}, K.~Damanakis\cmsorcid{0000-0001-5389-2872}, M.~Dragicevic\cmsorcid{0000-0003-1967-6783}, A.~Escalante~Del~Valle\cmsorcid{0000-0002-9702-6359}, R.~Fr\"{u}hwirth\cmsAuthorMark{2}\cmsorcid{0000-0002-0054-3369}, M.~Jeitler\cmsAuthorMark{2}\cmsorcid{0000-0002-5141-9560}, N.~Krammer\cmsorcid{0000-0002-0548-0985}, L.~Lechner\cmsorcid{0000-0002-3065-1141}, D.~Liko\cmsorcid{0000-0002-3380-473X}, I.~Mikulec\cmsorcid{0000-0003-0385-2746}, P.~Paulitsch, F.M.~Pitters, J.~Schieck\cmsAuthorMark{2}\cmsorcid{0000-0002-1058-8093}, R.~Sch\"{o}fbeck\cmsorcid{0000-0002-2332-8784}, D.~Schwarz\cmsorcid{0000-0002-3821-7331}, S.~Templ\cmsorcid{0000-0003-3137-5692}, W.~Waltenberger\cmsorcid{0000-0002-6215-7228}, C.-E.~Wulz\cmsAuthorMark{2}\cmsorcid{0000-0001-9226-5812}
% ...
\par}
\vskip\cmsinstskip
\dag:~Deceased\\
$^{1}$Also at Yerevan State University, Yerevan, Armenia\\
$^{2}$Also at TU Wien, Vienna, Austria\\
% ...
